\documentclass{aastex631}

\usepackage{color} %delete before submission
\usepackage{soul} %delete before submission

\shortauthors{Ruan, Keppens et al.}
\graphicspath{{./}{figures/}}
\begin{document}

\title{The Lorentz force at work: multi-phase magnetohydrodynamics throughout a flare lifespan}

\correspondingauthor{Wenzhi Ruan, Rony Keppens}
\email{ruanw@mps.mpg.de, rony.keppens@kuleuven.be}

\author[0000-0001-5045-827X]{Wenzhi Ruan}
\affiliation{Centre for mathematical Plasma Astrophysics, Department of Mathematics, \\
KU Leuven, Celestijnenlaan 200B, 3001 Leuven, Belgium}
\affiliation{Max Planck Institute for Solar System Research, \\ 
Justus-von-Liebig-Weg 3, 37077 Gottingen, Germany}

\author{Rony Keppens}
\affiliation{Centre for mathematical Plasma Astrophysics, Department of Mathematics, \\
KU Leuven, Celestijnenlaan 200B, 3001 Leuven, Belgium}

\author{Limei Yan}
\affiliation{Key Laboratory of Earth and Planetary Physics, Institute of Geology and Geophysics, \\
Chinese Academy of Sciences, Beijing, People's Republic of China}

\author{Patrick Antolin}
\affiliation{Department of Mathematics, Physics and Electrical Engineering, \\
Northumbria University, Newcastle upon Tyne, NE1 8ST, UK}

\accepted{Mar 27, 2024}
%% Command to document which AAS Journal the manuscript was submitted to.
%% Adds "Submitted to " the argument.
\submitjournal{ApJ}

%% Note that the \and command from previous versions of AASTeX is now
%% depreciated in this version as it is no longer necessary. AASTeX 
%% automatically takes care of all commas and "and"s between authors names.

%% AASTeX 6.31 has the new \collaboration and \nocollaboration commands to
%% provide the collaboration status of a group of authors. These commands 
%% can be used either before or after the list of corresponding authors. The
%% argument for \collaboration is the collaboration identifier. Authors are
%% encouraged to surround collaboration identifiers with ()s. The 
%% \nocollaboration command takes no argument and exists to indicate that
%% the nearby authors are not part of surrounding collaborations.

%% Mark off the abstract in the ``abstract'' environment. 
\begin{abstract}
The hour-long, gradual phase of solar flares is well-observed across the electromagnetic spectrum, demonstrating many multi-phase aspects, where cold condensations form within the heated post-flare system, but a complete three-dimensional (3D) model is lacking. Using a state-of-the-art 3D magnetohydrodynamic simulation, we identify the key role played by the Lorentz force through the entire flare lifespan, and show that slow variations in the post-flare magnetic field achieve the bulk of the energy release. Synthetic images in multiple passbands closely match flare observations, and we quantify the role of conductive, radiative and Lorentz force work contributions from flare onset to decay. This highlights how the non-force-free nature of the magnetic topology is crucial to trigger Rayleigh-Taylor dynamics, observed as waving coronal rays in extreme ultraviolet observations. Our C-class solar flare reproduces multi-phase aspects such as post-flare coronal rain. In agreement with observations, we find strands of cooler plasma forming spontaneously by catastrophic cooling, leading to cool plasma draining down the post-flare loops. As there is force balance between magnetic pressure and tension and the plasma pressure in gradual-phase flare loops, this has potential for coronal seismology to decipher the magnetic field strength variation from observations.

\end{abstract}

%% Keywords should appear after the \end{abstract} command. 
%% The AAS Journals now uses Unified Astronomy Thesaurus concepts:
%% https://astrothesaurus.org
%% You will be asked to selected these concepts during the submission process
%% but this old "keyword" functionality is maintained in case authors want
%% to include these concepts in their preprints.

%% From the front matter, we move on to the body of the paper.
%% Sections are demarcated by \section and \subsection, respectively.
%% Observe the use of the LaTeX \label
%% command after the \subsection to give a symbolic KEY to the
%% subsection for cross-referencing in a \ref command.
%% You can use LaTeX's \ref and \label commands to keep track of
%% cross-references to sections, equations, tables, and figures.
%% That way, if you change the order of any elements, LaTeX will
%% automatically renumber them.
%%
%% We recommend that authors also use the natbib \citep
%% and \citet commands to identify citations.  The citations are
%% tied to the reference list via symbolic KEYs. The KEY corresponds
%% to the KEY in the \bibitem in the reference list below. 

\section{Introduction} \label{sec:intro}

Flares occur frequently in solar and stellar atmospheres, and signal energetic events in association with compact objects and black holes of all sizes (e.g. \citealp{Shibata2011LRSP,Gunther2020AJ,Aliu2012ApJ}). The sudden brightenings seen across X-ray (or even more energetic) wavebands result from rapid release of magnetic energy through magnetic reconnection, where detailed kinetic plasma processes boost photon emissions and generate near-light-speed particles \citep{Aschwanden2017ApJ,Pontin2022LRSP}. The bulk of the plasma radiates in extreme ultraviolet (EUV) and soft X-ray (SXR) channels, and this thermal plasma emission can be enhanced by several orders of magnitude when a solar flare occurs. Solar flares dramatically impact the ionosphere of the Earth and can harm radio communication and navigation systems \citep{Tsurutani2009RaSc}. The study of solar flares led to the standard flare scenario based on lightcurve variations and detailed spatio-temporally resolved images. This standard flare model has been modeled frequently in 2D magnetohydrodynamic settings where a vertical current sheet is evolved to a flare loop configuration, and is recently revisited with a number of advanced 3D simulations~\citep{Shen2022NatAs,Ruan2023ApJ,Wang2023}, as well as with 2.5D models where self-consistent two-way coupling between electron beams and the multi-dimensional MHD scenario is incorporated~\citep{Ruan2020ApJ,Druett2023}, or where post-flare coronal rain is obtained in multi-D settings~\citep{Ruan2021ApJ}. It should be noted that these models deliberately adopt a simple initial magnetic topology, and as yet do not involve the actual full flux rope eruption process, but rather concentrate on the realistic thermodynamic and energetic evolution in and below the current sheet region. This should be contrasted with the simulations (in zero-beta settings, i.e. without thermodynamics) of actual eruptions that have provided insight into the evolving topological evolutions where an initial unstable flux rope is present \citep{Aulanier2012}. Such eruptive models, extended to full thermodynamics and concentrating on the current sheet fine-structure~\citep{Ye2023}, are currently feasible thanks to adaptive mesh refinement strategies provided by our open-source code MPI-AMRVAC~\citep{Keppens2023}.

A typical solar flare has two phases: the impulsive phase lasting few minutes and the gradual phase, lasting up to several hours \citep{Fletcher2011SSRv,Hudson2011SSRv,Benz2017LRSP}. Magnetic reconnection rapidly forms dense (electron number density $N_e \sim 10^{10}~\mathrm{cm}^{-3}$) and hot (temperature $T \sim 10~\mathrm{MK}$) coronal flare loop systems at the impulsive phase, during which a large amount of high energy electrons are generated and the electrons produce strong hard X-ray emission through the Bremsstrahlung mechanism. The density and temperature slowly return to pre-flare levels ($N_e \sim 10^{9}~\mathrm{cm}^{-3}, T \sim 1~\mathrm{MK}$) in the long-duration gradual phase. 
Although most of the EUV and SXR radiation is released during this extended gradual phase, we still know very little about it, but remote observations across EUV and SXR wavebands present us with many dynamically evolving aspects, revealing clear multi-phase, multi-thermal plasma evolutions. A recent spectroscopic study of coronal rain formation in an X2 solar flare provides new clues on how cool condensations form in the hot flare system and then dynamically evolve \citep{Brooks2024}. Here, we aim to model this intruiging aspect of the long-term flare evolution in a self-consistent manner.
Recent radiative MHD models from photosphere to corona, where interacting sunspot regions are forced to pass one-another, also demonstrate flares when the sunspots have close encounters, with indications of rain forming \citep{Rempel2023}, but our current standard flare setup allows for a clear cause-and-effect study that is fully reproducable and more easily analyzed.

During the gradual phase, flare loops are thought to gradually cool and become denser due to an interplay of thermal conduction and radiative losses, and eventually cold coronal rain forms in the cooling, dense postflare loops and this rain drains along the field to the solar surface (e.g. \citealp{Schmieder1995SoPh,Mason2022ApJ}). 
Recent field-aligned flare simulations show that coronal rain cannot form in flare loop systems when the magnetic field configuration does not vary, which indicates that multi-dimensional effects are crucial for gradual-phase flare development \citep{Reep2020ApJ,Reep2022ApJ}. Since the duration of the gradual phase can be several hours (e.g. \citealp{Reep2019ApJ}), with a cooling rate lower than theoretically predicted (e.g. \citealp{Bian2016ApJb}), additional heating mechanisms or factors inhibiting cooling may be present during the gradual phase. On EUV and SXR images during the impulsive and the gradual phase, a fan of bright spike-like rays is frequently seen above flare loops, and these rays demonstrate waving (e.g. \citealp{McKenzie1999ApJ,Khan2007A&A}). The coronal rays are thought to be associated with the so-called supra-arcade downflows (SADs) phenomenon, which represents downward-moving dark blobs that are often seen between the bright rays and are interpreted as sub-Alfv\'enic downflows (e.g. \citealp{McKenzie2009ApJ,Savage2011ApJ,Warren2011ApJ,Reeves2017ApJ,Samanta2021Innov,Xie2022ApJ,Tan2023MNRAS}). Waving coronal rays suggest that gradual flare loop systems are quite dynamic, pointing towards three-dimensional magnetohydrodynamic processes at play.

We present a three-dimensional (3D) magnetohydrodynamic simulation of a solar (or stellar) flare event which covers both the impulsive and the gradual phases. Our simulation captures many phenomena seen during the gradual phase, including coronal rain and waving coronal rays. We show that both rain and waving rays form spontaneously under the action of the Lorentz force, in combination with radiative losses allowing thermal instability. The energy stored in the non-force-free magnetic field of the flare loop system is slowly released through work done by the Lorentz force in the gradual phase, a process that has the potential to extend the duration of the flare. Our paper is organised as follows. We discuss our model ingredients in Sect.~\ref{sec:model}. All our results and analysis  are performed in Sect.~\ref{sec:Resu}. We conclude in Sect.~\ref{sec:summ}.

\section{Numerical model} \label{sec:model}

\subsection{Governing equations and open-source software used}

The MHD simulation is carried out using the open-source MPI-AMRVAC code \citep{Keppens2023}. The evolution of a solar flare is simulated in a square box with a domain of $-50~\mathrm{Mm} \leq x \leq 50~\mathrm{Mm}$, $-50~\mathrm{Mm} \leq y \leq 50~\mathrm{Mm}$ and $0 \leq z \leq 100~\mathrm{Mm}$. Here $z$ is the vertical direction and the $x-y$ plane is parallel to the solar surface. An equivalent resolution of $512 \times 512 \times 512$ is achieved with a block-adaptive mesh, where a 4-level adaptive refinement starts with a level-1 mesh of resolution $64 \times 64 \times 64$.
The cells in the flare loop systems and the reconnection current sheet have a size of $195~\mathrm{km} \times 195~\mathrm{km} \times 195~\mathrm{km}$.
The contributions of thermal conduction, radiative loss, and gravity have been taken into account. The governing equations are given by
\begin{eqnarray}
\frac{\partial \rho}{\partial t} + \nabla \cdot (\rho \mathbf{v}) & = &  0\,, \label{EqMHDrho} \\
\frac{\partial \rho \mathbf{v}}{\partial t} + \nabla \cdot (\rho \mathbf{v} \mathbf{v} + p_{\mathrm{tot}} \mathbf{I} - \mathbf{B}\mathbf{B}) & = &  \rho \mathbf{g}\,, \label{EqMHDmom} \\
\frac{\partial e}{\partial t} + \nabla \cdot (e \mathbf{v} + p_{\rm{tot}} \mathbf{v} - \mathbf{BB} \cdot \mathbf{v}) & = &  \rho \mathbf{g} \cdot \mathbf{v} +\nabla \cdot (\mathbf{\kappa} \cdot \nabla T)  \nonumber \\
& & + \nabla \cdot (\mathbf{B} \times \eta \mathbf{J}) -  Q_r + H_b\,, \label{EqMHDe} \\
\frac{\partial \mathbf{B}}{\partial t} + \nabla \cdot (\mathbf{v B} - \mathbf{B v}) & = & - \nabla \times (\eta \mathbf{J}) \,, \label{EqMHDB}
\end{eqnarray}
where $\rho$, $\mathbf{v}$, $\mathbf{B}$, $e$, $p_{\rm{tot}}$, $T$, $\mathbf{J}$ are density, velocity vector, magnetic field vector, energy  density (sum of kinetic energy, thermal energy and magnetic energy), total pressure (sum of thermal pressure and magnetic pressure), temperature, and current density vector, respectively. The gravity acceleration is given by $\mathbf{g} = -274 R_{\rm{s}}^2 / (R_{\rm{s}} + z)^2 \mathbf{\hat{z}} \ \rm m\ s^{-2}$, where $R_{\rm{s}}=696.1$ Mm is the radius of the sun. A thermal conductivity tensor $\mathbf{\kappa}=  \kappa_{\parallel} T^{5/2} \mathbf{\hat{b}}\mathbf{\hat{b}}$ is employed, where $\kappa_{\parallel} = 8 \times 10^{-7} \ \rm {erg\ cm^{-1}\ s^{-1}\ K^{-7/2}}$ and $\mathbf{\hat{b}}$ is a unit-tangent to the magnetic field vector. Saturation of thermal flux is employed \citep{Xia2018ApJS}. The optically thin radiative loss is given by
\begin{equation}
Q_r=N_e N_i \Lambda(T) ,
\end{equation}
where $N_e$ is electron number density, $N_i$ is ion number density and $\Lambda(T)$ is an optically thin cooling curve from \citet{Colgan2008ApJ}. Where the height is greater than 6 Mm, an extension of the cooling curve at low temperatures is activated (see curve `Colgan\_DM' in \citet{Hermans2021A&A}), allowing the coronal rain plasma to cool to $10,000~\mathrm{K}$ or below. In this simulation, we assume that the number ratio of hydrogen to helium is $1:0.1$, so we have $N_i=1.1N_H$ and $N_e=1.2 N_H$, where $N_H$ is the number density of hydrogen (protons). A background heating $H_b$ is adopted to compensate for the radiative loss outside the flaring region. The background heating is much weaker than the radiative loss, thermal conduction, or Lorentz work and therefore does not play an important role in the development of the flare loop systems, as will be clear from our detailed analysis. This background heating is necessary to sustain the corona against the radiative losses, in the region away from the flare site where we settle on a stationary stratified chromosphere to corona. Ohmic dissipation is included to reproduce the magnetic reconnection process, where $\eta$ is the resistivity. 

In a finite volume approach, the conservative form of MHD equations (Eqn.~(\ref{EqMHDrho}) - (\ref{EqMHDB})) is numerically solved using the classical `HLL' Riemann solver (initials of authors Harten, Lax and van Leer in \citet{Harten1983upstream}). Flux limiters are employed during the calculation of cell fluxes to avoid unphysical oscillations near high gradient regions. The robust second-order limiter `vanleer' (from \citet{vanLeer1974JCoPh}) is employed at the flaring regions and in the low atmosphere (the local mesh has a refinement level greater than three there). The high-precision third-order limiter `cada3' (from first author of \citet{Cada2009JCoPh}) is employed in the corona outside the flaring region (where the local mesh refinement level is less than or equal to three). The field-line-based transition region adaptive conduction (TRAC) method is employed for the low atmosphere (below height $h\leq 5~\mathrm{Mm}$) in our simulation to ensure that coronal density and temperature evolve correctly \citep{Johnston2020A&A,Zhou2021A&A}.

\subsection{Initial conditions}

\begin{figure}[h]
\includegraphics[width=\textwidth]{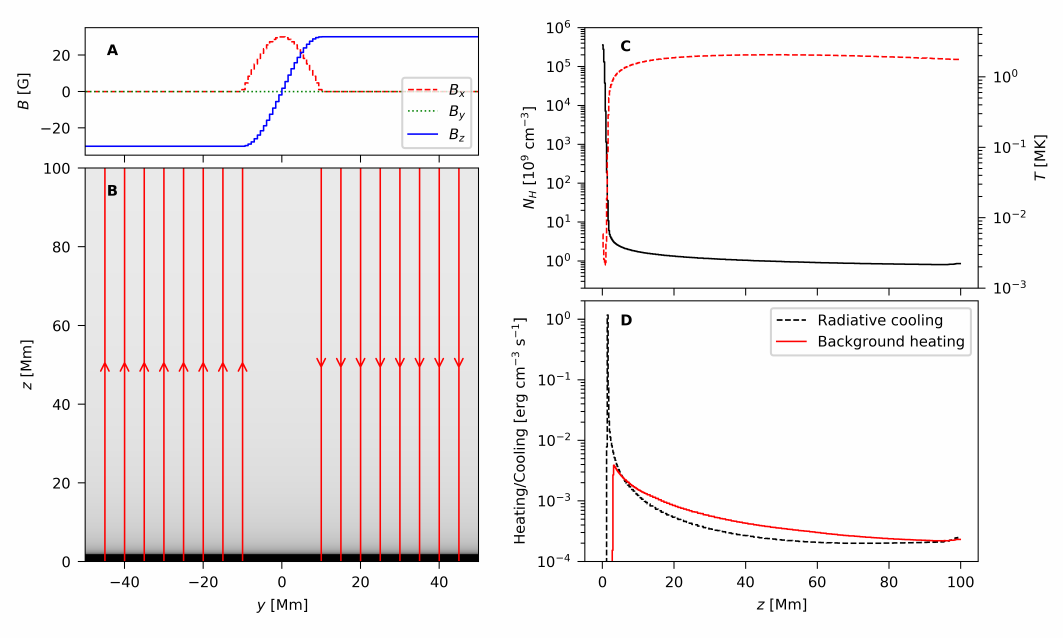}
\caption{ Initial conditions of the simulation. The initial magnetic field lines are displayed in panel (B), where the background is density. Panel (A) shows how the magnetic field components change in the $y$-direction. The variations of proton number density (black solid line) and plasma temperature (red dashed line) in the vertical $z$-direction are given by panel (C). The time-invariant background heating is demonstrated in panel (D), where the initial profile of radiative loss is also given.
\label{FigS1}}
\end{figure}

Our model covers the chromosphere, the transition region, and the corona. The initial density and temperature profiles are obtained from a 2.5D atmosphere relaxation simulation with a uniform upward magnetic field, in which the C7 temperature profile from \citet{Avrett2008ApJS}, a density profile calculated based on hydrostatic equilibrium, and an adaptive background heating are employed. More details on the relaxation are available in \citet{Ruan2021ApJ} and the results are demonstrated in Fig.~\ref{FigS1}. The background heating shown in Fig.~\ref{FigS1} is our source term $H_b$. The initial magnetic field is also quantified in Fig.~\ref{FigS1}. The initial magnetic field is a force-free field given by the following equations:
\begin{eqnarray}
B_x &=& \sqrt{B_0^2-B_z^2}, \label{EqBx}  \\
B_y &=& 0, \label{EqBy} \\
B_z &=& \left\{
	\begin{array}{ccc}
   	&-B_0, & y<-\lambda  \\
   	&B_0,  & y>\lambda   \\
   	&B_0 \sin [\pi y/(2\lambda)],  & {\mathrm{elsewhere}}\,, 
	\end{array}
	\right. \label{EqBz}
\end{eqnarray}
where $B_0=30~\mathrm{G}$ and $\lambda=10~\mathrm{Mm}$. The magnetic field reverses at the plane $y=0$, which allows for magnetic reconnection. 

\subsection{Resistivity prescription}

A time-varying resistivity is used in our simulation, which is given by
\begin{equation}
\eta (y,z;t) = 
	\left\{
		\begin{array}{ccc}
		  & \eta_1  \exp(- y^2/w_{\eta y}^2) \exp[- (z - h_{\eta})^2/w_{\eta z}^2] , & t < 450\ \mathrm{s}    \,, \\
		  & \eta_2 , & t \geq 450 \ \mathrm{s}  \,,
		\end{array}
	\right. \label{Eqeta}
\end{equation}
where $\eta_1=10^{-2}$, $\eta_2=10^{-3}$, $w_{\eta y}=10~\mathrm{Mm}$,  $w_{\eta z}=15~\mathrm{Mm}$, and $h_{\eta}=30~\mathrm{Mm}$. A localized resistivity at the first stage $t < 450~\mathrm{s}$ dissipates the wide initial current sheet, resulting in the formation of closed arcades at the low atmosphere below $h_{\eta}$ and a thin current sheet with a large magnetic field gradient in higher atmospheric regions. Thereafter, sustained magnetic reconnection occurs at the thinned current sheet, where we can adopt a uniform, low value of resistivity $\eta_2$ and let the reconnection evolve by a combination of fully resolved and numerical contributions. 

Anomalous resistivity, where a critical current threshold activates resistivity much stronger than that from Spitzer's theory, is often used in magnetic reconnection simulations to reproduce fast reconnection (e.g. \citealp{Yokoyama2001ApJ,Ruan2020ApJ}). In magnetic reconnection, the physics behind this anomalous resistivity is that microscopic instabilities can lead to this resistivity model \citep{Yokoyama2001ApJ}. It is still unclear whether this also works outside the reconnecting current sheet, such as in flare loops. In this work, we focus on flare loop systems in their gradual phase, rather than on reconnection in the current sheet. We therefore do not use anomalous resistivity after the initial phase of 450 seconds in this simulation, avoiding strong heating in the flare loop systems caused by such anomalous resistivity. Instead, we employed a relatively weak uniform resistivity in the gradual phase. In the simulation period $t<450~\mathrm{s}$, our adopted resistivity is only spatially located, and its magnitude $\eta \sim 10^{-2}$ obviously dominates the reconnection process. Given our effective resolution, our effective numerical resistivity can be estimated to be in the order of magnitude of $10^{-3}$ or lower. We employ an explicit uniform resistivity $\eta=10^{-3}$ in the period $t \geq 450~\mathrm{s}$ to control the overall numerical resistivity at play (combined explicit as source terms and numerical).

\subsection{Boundary conditions}

Periodic conditions are employed at the $x$-boundaries. 
Symmetric conditions are used for density, $v_x$, $v_z$, pressure, $B_x$ and $B_z$ at the sideways $y$-boundaries, and anti-symmetric conditions are employed for $v_y$ and $B_y$ at $y$-boundaries.
Here, a symmetric condition indicates $\Gamma(s)=\Gamma(-s)$, while anti-symmetric implies $\Gamma(s)=-\Gamma(-s)$, for scalar $\Gamma$ and where $s=0$ is at the boundary. Fixed conditions are employed at the bottom boundary (at $z=0$): the initial temperature at the lower boundary ghost cells is extrapolated linearly from the C7 temperature profile; the density is obtained based on hydrostatic equilibrium; the velocity is set to zero; and the magnetic field is given by Eqn.~(\ref{EqBx}) to (\ref{EqBz}). Zero gradient extrapolation is adopted for density at the upper boundary. Equal gradient extrapolation is employed for thermal pressure in the upper boundary ghost cells, where $p/(\mathrm{d}p/\mathrm{d}h)=50~\mathrm{Mm}$ as function of height $h=z$. Zero gradient extrapolation is employed for velocity and magnetic field components at the period $t<540~\mathrm{s}$, which leads to an open boundary. The velocity and magnetic field in the upper boundary ghost cells are treated differently from $t=540~\mathrm{s}$ onwards, when we switch to an effectively closed top boundary prescription: velocity is set to zero, a symmetric condition is employed for $B_x$ and $B_z$, and the anti-symmetric condition is employed for $B_y$.
This time-specific modification of the upper boundary is employed to limit the otherwise unrealistic infinite vertical expansion of the flare loop system. If the boundary conditions are not changed, the reconnection magnetic energy release rate will gradually decrease as the magnetic field strength and topology of the current sheet change. However, the expansion of the flare loops would not stop when they reach our upper boundary. The closed upper boundary forces the current sheet to represent a structure that is narrow in the middle and wider at both ends. In this case, due to the bending of the magnetic field, there will be an outward Lorentz force at the reconnection inflow region, thus suppressing the reconnection process and flare loop expansion. In real flares, the existence and associated expulsion of a flux rope will naturally introduce such effects.

\subsection{Observational reference data and forward modeling} \label{sec:obser}

We compare our simulated flare evolution in various passbands, and use selected reference data from (different) flare events to compare with: to that effect we also produce synthetic views on our numerical flare evolution. We will do so for EUV and H$\alpha$ images.

\begin{figure}[h]
\includegraphics[width=\textwidth]{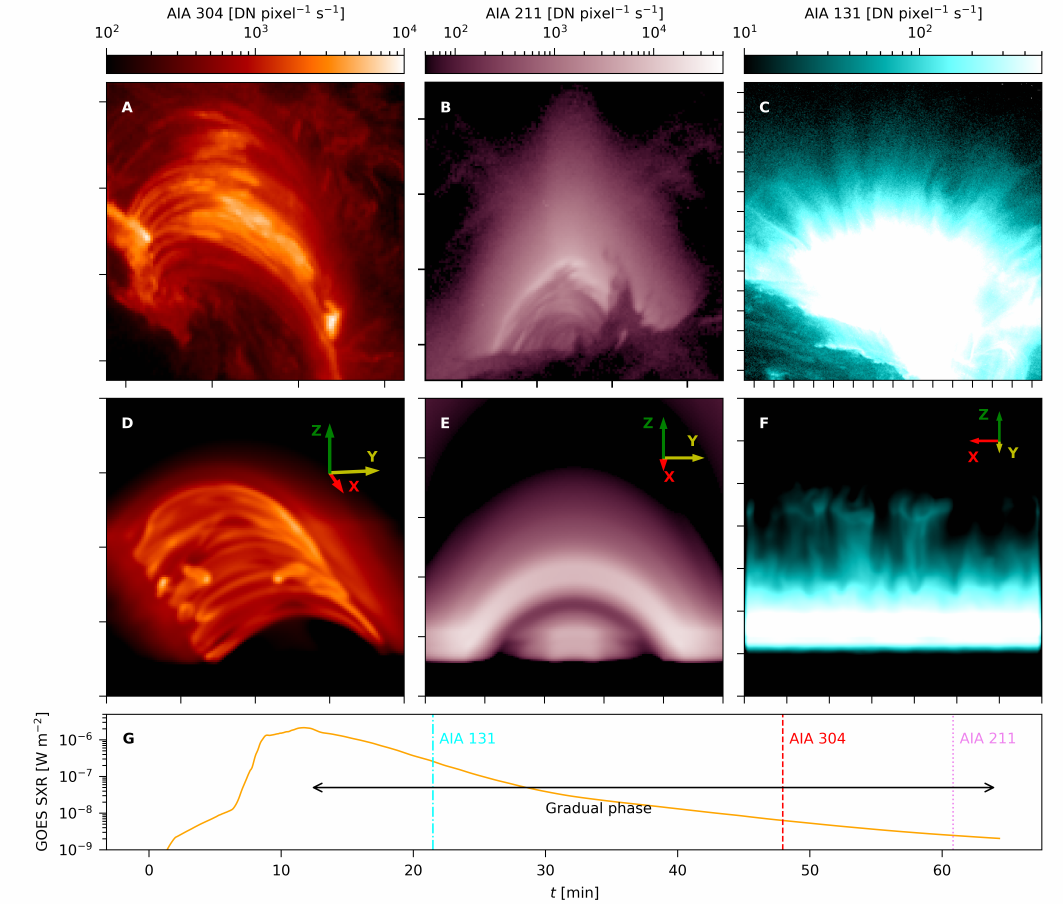}
\caption{ Solar flare loop systems at multiple passbands and multiple viewing angles. (\textbf{A}) An M6.5 flare observed at EUV 304 {\AA} wavelength by the Atmospheric Imaging Assembly (AIA) onboard the Solar Dynamics Observatory (SDO) \citep{Lemen2012SoPh}. This wavelength is sensitive to plasma temperature of $0.05~\mathrm{MK}$. This case has been reported in \citet{Mason2022ApJ}. (\textbf{B}) An X8.2 flare observed at EUV 211 {\AA} wavelength by AIA. The EUV 211 {\AA} wavelength is sensitive to plasma temperature of $2~\mathrm{MK}$. This case has been reported in \citet{Martinez2022ApJ}. (\textbf{C}) An M5.0 flare observed at EUV 131 {\AA} wavelength by AIA. This wavelength is sensitive to plasma temperatures of 0.4 and $10~\mathrm{MK}$. This case has been reported in \citet{Tan2023MNRAS}. (\textbf{D}) Synthetic EUV 304 {\AA} image from our simulation. (\textbf{E}) Synthetic EUV 211 {\AA} image from our simulation. (\textbf{F}) Synthetic EUV 131 {\AA} image from our simulation. (\textbf{G}) Synthetic GOES SXR 1-8 {\AA} flux from our simulation. In the AIA images, the distance between the axis tickmarks is 20 arcsec (14.5~Mm). Flare-driven coronal rains and fan of coronal rays are phenomena frequently observed in C, M, and X-class flares (e.g., \citealp{Khan2007A&A,Mason2022ApJ}). An animation of this figure (Movie Fig.~2) is provided. The movie has a real-time duration of 29 s. The period 0 to 10 s of the movie shows the synthetic 131 {\AA} view as well as 193 {\AA} view (not shown in the figure) from simulation time 14 min to 21 min. The direction of the LOS is rotated after the time reaches 21 min. The period 10 to 29 s of the movie shows the synthetic 304 {\AA} view from simulation time 35 min to 57 min. The direction of the LOS is rotated after the time reaches 57 min.
%Animations of this figure (Movie Fig2A and Fig2B) are provided. Movie Fig2A shows the synthetic 304 {\AA} view from simulation time 35 min to 57 min (real-time duration 19 s). Movie Fig2B shows the synthetic 131 {\AA} view as well as 193 {\AA} view from simulation time 14 min to 21 min (real-time duration 10 s).
\label{Fig1}}
\end{figure}

Observations of solar flare loop systems in EUV passbands (Fig.\ref{Fig1}A-C) are obtained by SDO/AIA \citep{Lemen2012SoPh}. 
The AIA/EUV images have a cadence of 12~s and a spatial sampling of $0.6 ''$ /pixel.
The aia~prep routine in the Solar Software \citep{Freeland1998SoPh} is applied to align AIA images.
The AIA 304~\AA\, image in Fig.~\ref{Fig1}A was observed at 18:58:08 UT on June 22, 2015. The AIA 211~\AA\, image in Fig.\ref{Fig1}B was observed at 16:16:37 UT on September 10, 2017. The AIA 131~\AA\, image in Fig.\ref{Fig1}C was observed at 14:30:34 UT on May 22, 2013. We note that these different events also sample different flare classes, but their overall morphological evolution can be considered similar, as also evidenced through standard flare model efforts presented in \citet{Druett2024}.

The possibility to generate EUV emission synthesis has been included in the open-source version of the MPI-AMRVAC 3.0 code \citep{Keppens2023}. Corresponding functions from the CHIANTI atomic database (version 8) are employed in the synthesis \citep{DelZanna2015A&A}. Optically thin assumptions are employed. In the high-density lower atmosphere ($z<5~\mathrm{Mm}$) that deviates from the optically thin condition, the EUV emissivity is set to zero. A point spread function has been used to include scattering effects by the instrument, and the synthesized images have the same resolution as the observations. The method given in \citet{Pinto2015A&A} is employed in the calculation of GOES SXR flux.

When synthesizing an H$\alpha$ image, we assume that at a height of $5~\mathrm{Mm}$ there is H$\alpha$ radiation of unit intensity propagating along a supposed line of sight (LOS) direction. The H$\alpha$ light is assumed to be partially absorbed when it meets the cold rain materials, and the absorption rate is calculated with the method in \citet{Heinzel2015A&A}. The impact of plasma motion is neglected in this calculation. An observation of flare-driven coronal rain at $\mathrm{H}\alpha$ wavelengths is found in \cite{Jing2016NatSR}.

\section{Results} \label{sec:Resu}

\subsection{Overall evolution from flare onset to decay}

We adopt a standard 3D flare model setup that evolves a current-sheet to its impulsive phase, including its realistic thermodynamic stratification in solar atmospheric conditions. After about seven minutes, an induced fast reconnection sets in, and the flare goes into its impulsive phase. By the 12th minute, the thermal SXR emission reaches its maximum value (Fig.~\ref{Fig1}G). After that, the SXR flux gently drops and the flare moves into its gradual phase. The simulated flare is a C-class flare, according to the classification of solar flares based on the peak GOES SXR flux. Figure~\ref{Fig1} and the associated movies demonstrate that we recover multiple typical observational behaviour across EUV wavebands throughout the entire gradual phase. Note how the AIA 304 and 211 views are representative for the late gradual phase, where we have obvious condensations in direct agreement with the selected flare observations. In the AIA 131 view, taken in the early decay phase of the flare, our side-on view clearly shows coronal rays above the arcade system. Our animated views (Movie Fig2) show the temporal evolution of the flare loop system in 304 (from 10 to 29 s of the movie) and in 131 and 193 (0 to 10 s of the movie) wavebands, while also addressing the role played by the LOS, by rotating the scene.

Typical temperature, density, and magnetic field configurations in the gradual phase are displayed in Fig.~\ref{Fig2}. Magnetic reconnection occurs above the flare loop systems. While the loops at the outer layers are newly formed, those at the inner layers were generated before or during the early impulsive phase. Newly formed loops get filled with high-temperature and high-density evaporative flows, and their temperature gradually decreases due to heat conduction and radiative cooling. The loops become shorter during this cooling process (Movie Fig3). During the gradual phase, the reconnection rate is slow, and the rate of energy injection is lower than the rate of energy loss caused by heat conduction and radiative cooling (this will be shown in Section~\ref{sec:energy}), so the temperature of the coronal flare region generally decreases slowly. Simultaneously, the density of this region drops over time, due to the continuous loss of material from the corona to the chromosphere \citep{Bradshaw2005A&A}. 

\subsection{Non-force-freeness of the evolving arcades}\label{sec:force}

The newly formed magnetic arcades at the outer layer of the flare region are far from a force-free stage, and the Lorentz force ($\mathbf{J} \times \mathbf{B}$) is pointing inward. Under the influence of this downward Lorentz force, the arcades are forced to shrink, meanwhile compressing the plasma below, until the Lorentz force and the pressure gradient ($-\nabla p$) reach a force balance. This can be quantified by introducing suitably averaged force measures, where we acknowledge that our initial setup is invariant along the $x$-direction (i.e. identical in all $y-z$ planes): the magnetic field has initially mostly $z$ components, and only within $|y|<\lambda$ is the field showing out-of-plane $x$-components $B_x(y,t=0)$. All variations in the $x$ direction are due to spontaneously ensuing turbulent fluctuations, as also our initial anomalous resistivity is $y-z$-dependent only. After the initial 450 seconds, low-lying loops formed with again essentially purely $y-z$ variations. Therefore, we can meaningfully average out the $x$-variation at all times in our study, to get the essential force-balance achieved across the arcade. This is true throughout the 3D flare evolution, as our magnetic topology closely follows the cartoonized 2.5D standard flare model. Indeed, cross-sectional views at varying $x$ values at a fixed time look overall similar, while differing in details. Fig.~\ref{Fig3} shows the distributions of the mean Lorentz force, gravity, and pressure gradient force on the $y-z$ plane. Since an average in the $x$ direction is involved, we refer to these forces as mean forces. Averaging not only better shows the force distribution, but also smooths out the influence of the occuring turbulence. Here we give the details of the calculations for these mean forces.

The calculation of the mean Lorentz force is based on the mean magnetic field configuration. The mean magnetic field is given by
\begin{eqnarray}
\bar{B}_s (y,z) &=& \int_{x_{\mathrm{min}}}^{x_{\mathrm{max}}} B_s (x,y,z) \mathrm{d}x / (x_{\mathrm{max}} - x_{\mathrm{min}})\,, 
\end{eqnarray}
for field components along $s=x,y,z$, 
where $(x_{\mathrm{min}},x_{\mathrm{max}})$ is the domain in the $x$-direction. The vertical component of the mean Lorentz force shown in Fig.~\ref{Fig3}A is then given by
\begin{equation}
\bar{F}_{B,z} = \bar{J}_x \bar{B}_y - \bar{J}_y \bar{B}_x 
= (\frac{\partial \bar{B}_z}{\partial y} - \frac{\partial \bar{B}_y}{\partial z}) \bar{B}_y - \frac{\partial \bar{B}_x}{ \partial z} \bar{B}_x \,, 
\end{equation}
where $\bar{J}_x$ and $\bar{J}_y$ are current density components calculated from the mean magnetic field, and the $x$-variation of the mean magnetic field is zero (in accord with our statements above on the similarity across $x$).
The mean gravity shown in Fig.~\ref{Fig3}B is given by
\begin{equation}
\bar{F}_{g,z} (y,z) = - g(z) \int_{x_{\mathrm{min}}}^{x_{\mathrm{max}}} \rho (x,y,z) \mathrm{d}x / (x_{\mathrm{max}} - x_{\mathrm{min}})\,, 
\end{equation}
where $\rho$ is plasma density, $g(z) = 274 \, (z/R_{\mathrm{sun}})^2 ~\mathrm{m}~\mathrm{s}^{-2}$, and $R_{\mathrm{sun}}=696.1~\mathrm{Mm}$ is the solar radius. The mean pressure gradient force is calculated from the average thermal pressure, hence it is given by
\begin{equation}
\bar{p} (y,z) = \int_{x_{\mathrm{min}}}^{x_{\mathrm{max}}} p (x,y,z) \mathrm{d}x / (x_{\mathrm{max}} - x_{\mathrm{min}})\,.  \label{EqApth}
\end{equation}
The vertical component of the mean pressure gradient force shown in Fig.~\ref{Fig3}C is then simply
\begin{equation}
\bar{F}_{p,z} = - \frac{\partial \bar{p}}{ \partial z}\,. 
\end{equation}
Moreover, a line-based analysis is also shown in Fig.~\ref{Fig3}, where the black line in Fig.~\ref{Fig3}D is a `magnetic field line' derived from the average magnetic field $(\bar{B}_y,\bar{B}_z)$, and the blue line is a vertical line from $(y = 0, z = 3~\mathrm{Mm})$ to $(y = 0, z = 40~\mathrm{Mm})$. Fig.~\ref{Fig3}E shows the density distribution (black crosses) along this magnetic `field line', derived from the temperature curve and gravity stratification ($\rho \mathbf{g} - \nabla p = 0$). This computed (black crosses) density profile is given by
\begin{equation}
\rho^{\times} (s) = \frac{p^{\times}(s)}{R \bar{T}(s)} = \frac{1}{R \bar{T}(s)} \bar{p} (s_0) \exp[ \int_{s_0}^s \frac{\mathbf{g} \cdot \mathrm{d} \mathbf{l}}{R \bar{T}(l)} ] , \label{Eqrhos}
\end{equation}
where the symbol $^\times$ indicates our derived variable, $\mathbf{l}$ and $l$ is the integral pathlength (tangent of the `field line'), $s$ denotes  location along the path, where $s_0$ is the integral starting point located at $z=3~\mathrm{Mm}$ on the `field line'. Furthermore, $\mathbf{g} = - g \mathbf{e}_z$, $R$ is the universal gas constant and the average temperature is computed from
\begin{equation}
\bar{T} (y,z) =  \int_{x_{\mathrm{min}}}^{x_{\mathrm{max}}} p (x,y,z)~ \mathrm{d}x / \int_{x_{\mathrm{min}}}^{x_{\mathrm{max}}} \rho (x,y,z)  R~ \mathrm{d}x . 
\end{equation}
Eq.~(\ref{Eqrhos}) uses the following derivation of the pressure profile:
\begin{eqnarray}
\mathrm{d}p (s) = \rho (s) \mathbf{g} \cdot \mathrm{d} \mathbf{s} =  \frac{p(s)}{R T(s)} \mathbf{g} \cdot \mathrm{d} \mathbf{s},  \\
\int_l \frac{\mathrm{d} p(s)}{p(s)} = \int_l \frac{\mathbf{g} \cdot \mathrm{d} \mathbf{s}}{R T(s)},  \\
p(s) = p (s_0) \exp[ \int_{s_0}^s \frac{\mathbf{g} \cdot \mathrm{d} \mathbf{l}}{R T(l)} ]. 
\end{eqnarray}
The perpendicular `field line' density distribution (blue triangles) derived from the temperature curve and pressure gradient force-Lorentz force balance ($\mathbf{J} \times \mathbf{B} - \nabla p =0$) is also given in Fig.~\ref{Fig3}E. This predicted vertical (blue triangle) density profile is given by
\begin{eqnarray}
\rho^{\triangle} (s) = \frac{p^{\triangle}(s)}{R \bar{T(}s)} 
	&=& \frac{1}{R \bar{T}(s)} [\bar{p} (s_0) + \int_{s_0}^s \frac{\mathrm{d} p^{\triangle}(l)}{\mathrm{d} l} \mathrm{d}l] \nonumber \\
	&=& \frac{1}{R \bar{T}(s)} [\bar{p} (s_0) + \int_{s_0}^s \bar{F}_{B,z} (l) \mathrm{d}l], 
\end{eqnarray}
where the integral path is the blue vertical line in Fig.~\ref{Fig3}D, and symbol $^\triangle$ indicates our derived variable, where $s_0$ is the endpoint of the line located at $z=3~\mathrm{Mm}$.

Our movie accompanying Fig.~\ref{Fig3} quantifies these mean forces for 7 minutes starting after $t=14$ minutes. At all times, vertical Lorentz forces balance pressure gradients, so a non-force-free state is evidently realized throughout.
Indeed, our model shows that the pressure gradient and the Lorentz force in gradual-phase flare loops are closely balanced ($\mathbf{J} \times \mathbf{B} -\nabla p \approx 0$), while gravity is much weaker than these two forces (Fig.~\ref{Fig3}A-D and its Movie). Such a force balance controls the cross-magnetic-field density distribution (Fig.~\ref{Fig3}E), with typically higher density in the inner layers (also seen in Fig.~\ref{Fig2}B). 
The field-aligned density distribution is governed by a hydrostatic balance instead, namely that of gravity and pressure gradient forces (Fig.~\ref{Fig3}E). As a result, density along a loop tends to decrease with height. 

The cool and dense loops in the lower layers represent the ideal environment to form coronal rain. Radiative loss eventually dominates the cooling in these loops (see section~\ref{sec:energy}). Radiative-loss-driven catastrophic cooling initiates in the dense loops after they cooled to about $1~\mathrm{MK}$ \citep{Field1965ApJ,Cargill2013ApJ}. As a result, sudden condensations occur, creating colder ($\sim0.01~\mathrm{MK}$), denser rain blobs (Fig.~\ref{Fig2} and its movie). These rain blobs appear as bright spots in EUV 304 $\mathrm{\AA}$ images (Fig.~\ref{Fig1}A,D and its Movie Fig2A) and dark in H$\alpha$ images. 
In Fig.~\ref{FigS7} and its movie, we show how the development of the rain is clearly seen in isotemperature animations (showing regions with temperature lower than $25,000~\mathrm{K}$ and proton number density higher than $10^9~\mathrm{cm}^{-3}$). Fig.~\ref{FigS8} and its movie translates these simulations to an H$\alpha$ proxy, as explained in Section~\ref{sec:obser}. These views qualitatively agree with the stranded nature of the field-aligned post-flare rain condensation observed in many flare events, e.g. in the work by~\cite{Jing2016NatSR}.
Through the formation of coronal rain, the inner loops gradually become empty (see also Movie~Fig3), as suggested by current flare models \citep{Priest2002A&ARv}.

We note that there are other ways to calculate the mean Lorentz force, e.g. one could calculate the Lorentz force from the original current and magnetic field and then average the Lorentz force (Appendix \ref{sec:meanforce}). We verified that for our setting (typified by an invariant direction initially), these two methods qualitatively agree, although they obviously differ in details (especially where turbulence is active). The main force balance conclusions we noted are completely identical between both methods. The advantage of the method used in our article is that it can directly show the relationship between the average magnetic field and the average pressure. In flare observations, we may thus be able to infer the gradual-phase magnetic field based on this relationship.

\subsection{Energetic analysis}\label{sec:energy}

The delicate balance between Lorentz force and pressure gradient evolves dynamically during the gradual phase. Influenced by downward enthalpy fluxes, thermal conduction, and radiative losses, the pressure in the flare region is continuously decreasing \citep{Bradshaw2010ApJ}. The Lorentz force will thus temporarily be stronger than the pressure gradient, and the magnetic arcades move down, compressing the plasma and reaching a new equilibrium. The Lorentz force does work during this procedure, releasing the energy in the magnetic field and making up for the internal energy lost. Instead of being converted directly into internal energy (as is the case for Ohmic heating in reconnection regions), the postflare magnetic energy is first transformed into kinetic energy and then, by compression,  partially into internal energy. The Lorentz work, which is of the same magnitude as the radiative losses and thermal conduction (Fig.~\ref{Fig4}A,C), thereby helps to slow down the cooling of the flare region. Our C-class flare adopted a coronal magnetic field strength of $B_0 = 30~\mathrm{G}$, but in X-class flares (e.g. \citealp{Fleishman2020Sci}), magnetic field strengths can be an order of magnitude higher. The energy provided by the work done by the Lorentz force ($\sim B_0^2$) will then be two orders of magnitude higher and may effectively prolong the duration of the gradual phase. This is consistent with the finding that in X-class flares, the magnetic flux from the footpoints of flare loop systems has a positive correlation with flare durations \citep{Reep2019ApJ}. 
On the other hand, prolonging the duration of lower-class flares might be dominated by other reasons, such as on-going reconnection and long-lasting cooling (e.g. \citealp{Cargill1983ApJ,Liu2013ApJ,Li2014ApJ,Toriumi2017ApJ,Reep2017ApJ}), since the duration of small flares appears to be independent of the magnetic flux (also see \citet{Reep2019ApJ}). The way we quantify the detailed energy evolution is given below.

The distributions of Lorentz force work, radiative loss, and thermal conductive heating/cooling on the $y-z$ plane are shown in Fig.~\ref{Fig4}A,B. The method to calculate the dimensionally-reduced distribution of the energy conversion rates is as follows:
\begin{eqnarray}
\bar{F}_{\mathrm{Lorentz}} (y,z) &=& \int_{x_{\mathrm{min}}}^{x_{\mathrm{max}}} \frac{[\mathbf{J}(x,y,z) \times \mathbf{B}(x,y,z)] \cdot \mathbf{v}(x,y,z)}{(x_{\mathrm{max}} - x_{\mathrm{min}})} \mathrm{d}x ,  \\
\bar{F}_{\mathrm{Radiative}} (y,z) &=& \int_{x_{\mathrm{min}}}^{x_{\mathrm{max}}} \frac{Q_r (x,y,z)}{(x_{\mathrm{max}} - x_{\mathrm{min}})} \mathrm{d}x , \label{EqD22} \\
\bar{F}_{\mathrm{Conductive}} (y,z) &=& \int_{x_{\mathrm{min}}}^{x_{\mathrm{max}}} \frac{\nabla \cdot [\mathbf{\kappa} \cdot \nabla T(x,y,z)]} {(x_{\mathrm{max}} - x_{\mathrm{min}})} \mathrm{d}x . 
\end{eqnarray}
Note that the conductive heating or cooling rates given in Fig. \ref{Fig4}B are here estimated in post-processing, without accounting for specific numerical effects, like the heat flux limiter or cooling flux adjustments (TRAC aspects), as well as flux fixes active at the borders of AMR grids. Fig.~\ref{Fig4}C shows the instantaneous 3D-space-integrated energy release rates, and their calculation happens as follows:
\begin{eqnarray}
I_{\mathrm{Rec}} (t) &=& \int_{S2} \int_{x_{\mathrm{min}}}^{x_{\mathrm{max}}} \left( [\mathbf{J}(\mathbf{x},t) \times \mathbf{B}(\mathbf{x},t)] \cdot \mathbf{v}(\mathbf{x},t) + \eta J^2 (\mathbf{x},t) \right) ~\mathrm{d}x ~\mathrm{d}S, \\
I_{\mathrm{Lor}} (t) &=& \int_{S1} \int_{x_{\mathrm{min}}}^{x_{\mathrm{max}}} [\mathbf{J}(\mathbf{x},t) \times \mathbf{B}(\mathbf{x},t)] \cdot \mathbf{v}(\mathbf{x},t) ~\mathrm{d}x ~\mathrm{d}S, \\
I_{\mathrm{Rad}} (t) &=& \int_{S1} \int_{x_{\mathrm{min}}}^{x_{\mathrm{max}}} Q_r (\mathbf{x},t) ~\mathrm{d}x ~\mathrm{d}S,  \\
I_{\mathrm{Con}} (t) &=& \int_{S1} \int_{x_{\mathrm{min}}}^{x_{\mathrm{max}}} \nabla \cdot [\mathbf{\kappa} \cdot \nabla T (\mathbf{x},t)] ~\mathrm{d}x ~\mathrm{d}S, \\
 I_{\mathrm{Eva}} (t) &=& \int_{S1b} \left [\frac{1}{2} \rho (\mathbf{x},t) ~ v^2 (\mathbf{x},t) + \frac{p (\mathbf{x},t)}{\gamma-1} \right ] ~ \mathrm{max}[0,v_z (\mathbf{x},t)] ~ \mathrm{d}S,
\end{eqnarray}
where $S1$ indicates the 2D area where high-density flare loop systems are located in the $y-z$ plane (referring to the region inside the solid curves in panels (A)-(C) of Fig.~\ref{FigS2}), $S2$ indicates the area around the reconnection current sheet in the $y-z$ plane (referring to the region inside the dashed curves in panels (A)-(C) of Fig.~\ref{FigS2}), and $S1b$ (a 2D horizontal surface) indicates the lower border of the region $S1$, $Rec$ stands for magnetic reconnection, $Lor$ stands for Lorentz force work, $Rad$ stands for radiative loss, $Con$ stands for thermal conduction and $Eva$ stands for upward chromospheric evaporations. These time-varying integral regions $S1$ and $S2$ are shown in the Movie of Fig.~\ref{Fig4}.

\citet{Birn2009ApJ} emphasizes the role of the Lorentz force work in converting reconnection energy.
We also here briefly analyse the energy conversion in the combined system during the early impulsive phase, where it is known that an MHD approximation does not capture essential non-thermal aspects of the flare evolution~\citep{Ruan2020ApJ}. 
In our 3D MHD simulation, the Lorentz force work at the sideways bounding slow shocks in region $S2$ is primarily responsible for the energy release in magnetic reconnection, consistent with the original Petschek steady, fast reconnection model. In panels (B) and (C) of Fig.~\ref{FigS2}, respectively, the spatial distributions of the Lorentz force work and Ohmic heating at the impulsive phase are displayed. In panel (D) of Fig.~\ref{FigS2}, integrals of Lorentz force work and Ohmic heating in region $S2$ are compared.  The release of magnetic energy is much more effectively accomplished by the work done by the Lorentz force than by the Ohmic heating, as seen in Fig.~\ref{FigS2}, panels (B)-(D). Figure \ref{FigS2} shows that the effect of ohmic heating caused by the uniform resistivity is much weaker than that of Lorentz force work. The effect of this ohmic heating is negligible for the evolution of flare loop systems. 

For completeness, the energy evolution of the flare loop system in region $S1$ is analysed simultaneously. The evolution of the thermal energy in flare loop systems tracks the evolution of the energy released through reconnection very well during our impulsive phase, and the thermal energy is close to half of the released magnetic energy (Fig.~\ref{FigS2}E), as the other half would go into the (eventually ejected) fluxrope system above. The thermal energy shown is given by
\begin{equation}
E_{\mathrm{th}} (t) = (x_{\mathrm{max}} - x_{\mathrm{min}}) \int_{S1} \frac{\bar{p} (y,z,t)}{\gamma-1}  ~\mathrm{d}S \,, 
\end{equation}
and the released magnetic energy is quantified by
\begin{equation}
E_{\mathrm{rc}} (t) = \int_0^t I_{\mathrm{Rec}} (\tau) ~\mathrm{d} \tau \,. 
\end{equation}
This ratio is reasonable when considering that roughly half of the released energy is transported upward and leaves the simulation domain.
To further understand the intricate relationship between the energy release in the reconnection region and the energy evolution of the flare loop systems, a detailed analysis of energy conversion and transport can be performed (e.g., \citealp{Birn2009ApJ,Reeves2010ApJ}). However, our main effort here concentrates on the gradual phase, and especillay the multi-thermal aspects at play that drive condensation formations, which is why we emphasize the role played by the non-adiabatic effects included in our study: radiation and conduction.

We conclude that magnetic energy can be slowly released in the flare loops through Lorentz work during the gradual phase. However, the role that this released energy plays in extending the duration of the flare remains to be evaluated. More detailed analysis, where we also vary this energy budget across different classes of flares is left for future work.

\subsection{Instability analysis} \label{sec:Inst}

Although the Lorentz force and pressure gradient establish a dynamically evolving force balance inside the gradual phase flare loops, this mechanical equilibrium is actually (linearly) unstable in the flare region where it is fully dominated by these two forces. %Since the cross-field density distribution deviates from purely gravitational stratification (Fig.~\ref{Fig3}D and \ref{Fig3}E) and the magnetic field is not potential, there is a lot of free energy in the system. We find that Rayleigh-Taylor instability inevitably arises (Fig.~\ref{Fig5}E) and releases the free energy. 
We find that Rayleigh-Taylor instability inevitably arises (Fig.~\ref{Fig5}E), in an environment where the cross-field density distribution deviates from purely gravitational stratification (Fig.~\ref{Fig3}D and \ref{Fig3}E) and the magnetic field is not potential.
Previous studies invoked the high-speed reconnection outflow for inducing Rayleigh-Taylor instability in the impulsive phase and explaining the formation of SADs \citep{Guo2014ApJ,Shen2022NatAs}. 
Here we show instead that the occurrence of Rayleigh-Taylor instability in flare regions is unavoidable throughout the gradual phase, and does not require the participation of high speed reconnection outflows: the detailed 3D magnetic, density and pressure variations induced by the gradually evolving postflare field configuration inherently bring the entire region to Rayleigh-Taylor induced mixing. 
An uneven distribution of EUV emissivity reflects the instability effects on temperature and density distributions (Fig.~\ref{Fig5} and Movie~Fig9). Waving spike-like coronal rays appear in EUV images as a result (Fig.~\ref{Fig1}F and Movie~Fig2). 
In our simulation covering the entire postflare gradual phase, internal energy feeds the Rayleigh-Taylor instability, and a significant portion of the internal energy comes from the chromosphere in the impulsive phase through evaporations (Fig.~\ref{Fig4}C). Therefore, an energy transfer pathway of ``reconnection - fast electron deposition/thermal conduction - evaporations - coronal rays" is at play. In contrast, the main energy transfer path in \citet{Guo2014ApJ} and \citet{Shen2022NatAs} is stated as ``reconnection - fast outflows - SADs".

We make a brief analysis of the Rayleigh-Taylor process in what follows.
Flare loop systems generally have high temperatures and thus have a large scale height. Assuming a loop temperature of $T = 5~\mathrm{MK}$, then the corresponding scale height is
\begin{equation}
H = \frac{k_{\mathrm{B}} T}{m g} \approx 300~\mathrm{Mm},
\end{equation}
where $k_B = 1.38\times 10^{-16}~ \mathrm{cm^2~g~s^{-2}~K^{-1}}$ is Boltzmann constant, the mean mass of particle $m=8 \times 10^{-25}~\mathrm{g}$ is supposed to be half that of a proton, and gravity is supposed to be a constant $g=2.74\times10^4~ \mathrm{cm~s^{-2}}$. As a typical peak proton number density inside flare loops is $10^{10}~ \mathrm{cm^{-3}}$ and that above flare loops is $10^{9}~ \mathrm{cm^{-3}}$, the distance one estimates from a balance between gravity and pressure gradients in an isothermal setting for these density contrasts leads to a length estimate of
\begin{equation}
L = -H \ln (\frac{10^{9}}{10^{10}}) \approx 700~\mathrm{Mm}.
\end{equation}
Such a length is much larger than the size of flaring regions. Hence, the observed density contrasts on the smaller scale of the loop system must be maintained by means of the Lorentz force, where it balances a non-negligible pressure gradient. Although the force balance can be achieved by the Lorentz force, this balance is unstable and can easily lead to Rayleigh-Taylor instability.

Here we take Fig.~\ref{Fig5} as an example to briefly analyze the Rayleigh-Taylor instability in the gradual-phase flare loops. Fig.~\ref{Fig5}A shows the density distribution at slice $y=0$, and this distribution is shown more clearly in Fig.~\ref{FigS3}A. Finger-like structures are continually generated at a height of 20 Mm due to Rayleigh-Taylor instability (also see Movie Fig9).  From the velocity distribution in Fig.~\ref{FigS3}B, it can be seen that there is no high-speed flow of hundreds of $\mathrm{km~s^{-1}}$ near the roots of the structures, indicating that the formation of the structures is independent of SADs. The pressure gradient is large at this height (Fig.~\ref{FigS3}C,D), and the pressure gradient force promotes the upward growth of the structure. When there is a huge density difference on both sides of a interface, the growth rate of the Rayleigh-Taylor instability can be written simply as
\begin{equation}
\omega=\sqrt{k a},
\end{equation}
where $k$ is wavenumber, $a$ is acceleration and $a=-|g|$ in gravity dominated Rayleigh-Taylor instability. Here, the pressure gradient force takes the role of gravity and $a = \nabla p \cdot \mathbf{e}_z / \rho$, where $\rho$ is density. With a wave vector $\mathbf{k}$ perpendicular to the magnetic field, the magnetic field has little inhibitory effect on this instability. Suppose $k=1/(10~\mathrm{Mm})$, $\mathrm{d}p/ \mathrm{d}z = - 10^{-8}~ \mathrm{Ba~cm^{-1}}$ (refers to Fig.~\ref{FigS3}D) and $\rho=10^{-14}~\mathrm{g~cm^{-3}}$ ($N_{\mathrm{H}} \approx 6\times10^{9}~\mathrm{cm}^{-3}$), then an acceleration $a=-10^{6}~\mathrm{cm~s^{-2}}$ and a growth rate of $|\omega| \approx 1/30~\mathrm{s^{-1}}$ are obtained. The instability hence grows on a sub-minute timescale.

\section{Summary} \label{sec:summ}

In summary, our magnetohydrodynamic model captures many of the mysterious multi-phase aspects witnessed in multi-frequency views of solar flare events, especially those occurring in the extended, hour-long gradual phase. 
We showed that models which do not account for the work delivered by Lorentz forces (such as those assuming pure field-aligned processes) entirely miss the most important ingredient to explain the flare sustained and gradual energy release across EUV and SXR channels. 
We reproduce three-dimensional multi-thermal processes in unprecedented detail. We showed there is sustained force balance between the magnetic pressure and tension and the plasma pressure inside the slow-evolving gradual-phase flare loops. This could be used to figure out the magnetic field strength variation from EUV and SXR observations, if both plasma density and temperature can be deduced.
Since the Lorentz force scales quadratically with field strength, stellar flaring events should also demonstrate a clear correlation between their gradual phase duration and their magnetic field strength.

We argue theoretically and demonstrate with our simulation that flare loop systems are in turbulent states due to Rayleigh-Taylor instability, when the loop density remains high. The coronal rays observed in the gradual phase may be the external manifestation of the Rayleigh-Taylor instability. We suggest that sub-Alfv\'enic downflows (SADs) can lead to the generation of coronal rays, but not all coronal rays are related to such flows. It is foreseeable that transverse waves will be continuously excited in the turbulent loop systems and transport a large amount of energy from the corona along the loops to the underlying atmosphere, accelerating the cooling of the flare loops. On the other hand, the turbulence has the potential to reduce downward thermal conductive heat flux, which in turn slows down the loop cooling \citep{Bian2016ApJa,Bian2016ApJb,Emslie2018ApJ,Xie2023ApJ}. The impact of the turbulence on the flare loop systems development is still to be studied, as turbulence inherently requires even more extreme resolution simulations.

\begin{figure}
\includegraphics[width=\textwidth]{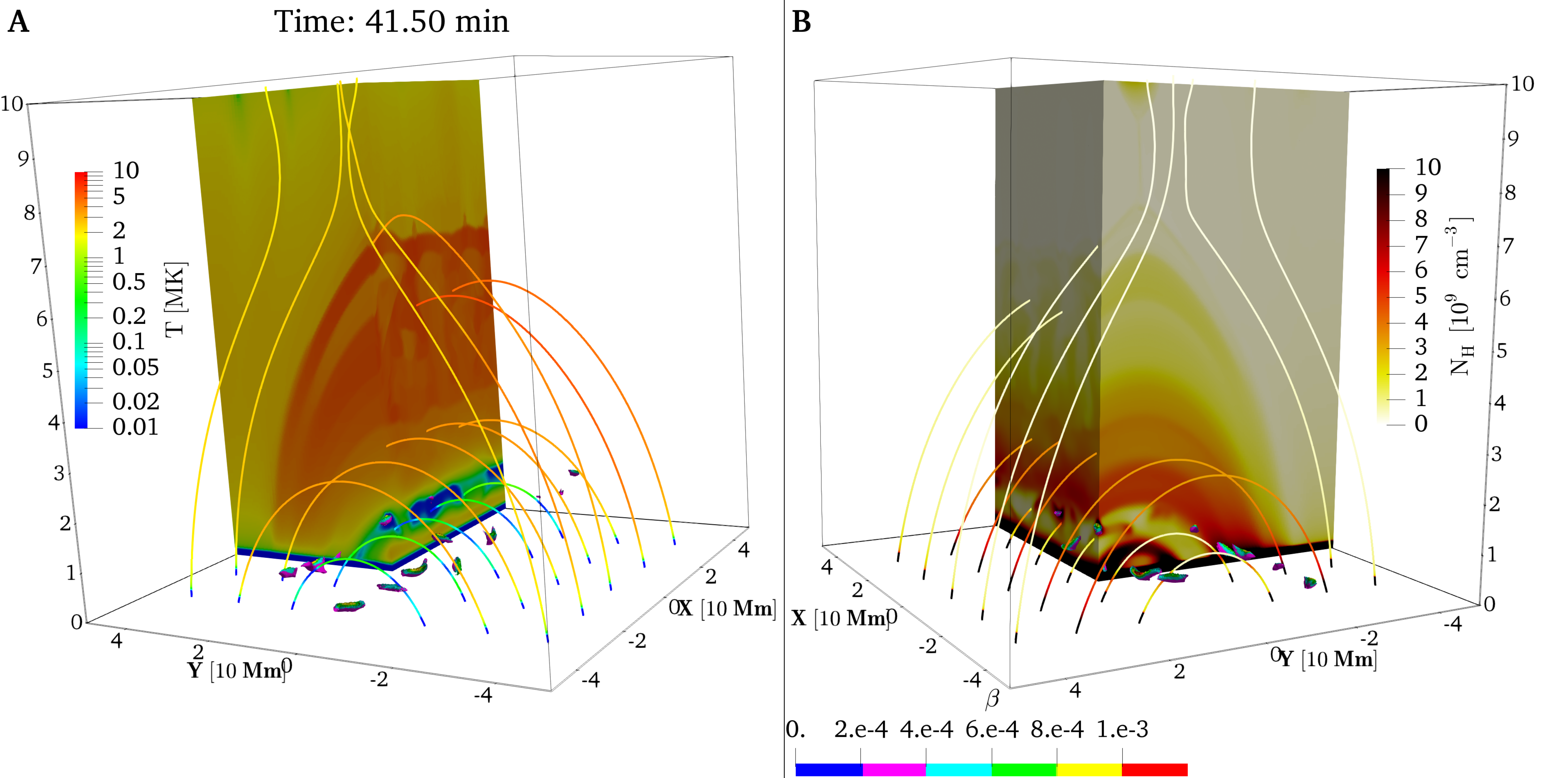}
\caption{ {Multi-phase aspects in the gradual-phase flare loop systems.} We show temperature distribution ($T$ in panel (A)), proton number density ($N_H$ in panel (B)), and magnetic field configuration (solid lines in both panels). The coloring of magnetic field lines is according to the local temperature (panel (A)) or density (panel (B)). Time is shown at the top of panel (A). Coronal rain blobs, which have proton number densities greater than $10^{10}~\mathrm{cm}^{-3}$ and temperatures less than $15,000~\mathrm{K}$, are visualized as isosurfaces seen in both panels. The surface coloring of these rain blobs gives the local plasma beta ($\beta$), which quantifies the ratio of thermal to magnetic pressure. An animation of this figure (Movie Fig3) is provided. The animation covers $\sim$19 minutes of physical time starting at t = 28 min (real-time duration 9 s).
\label{Fig2}}
\end{figure}

\begin{figure}
\includegraphics[width=\textwidth]{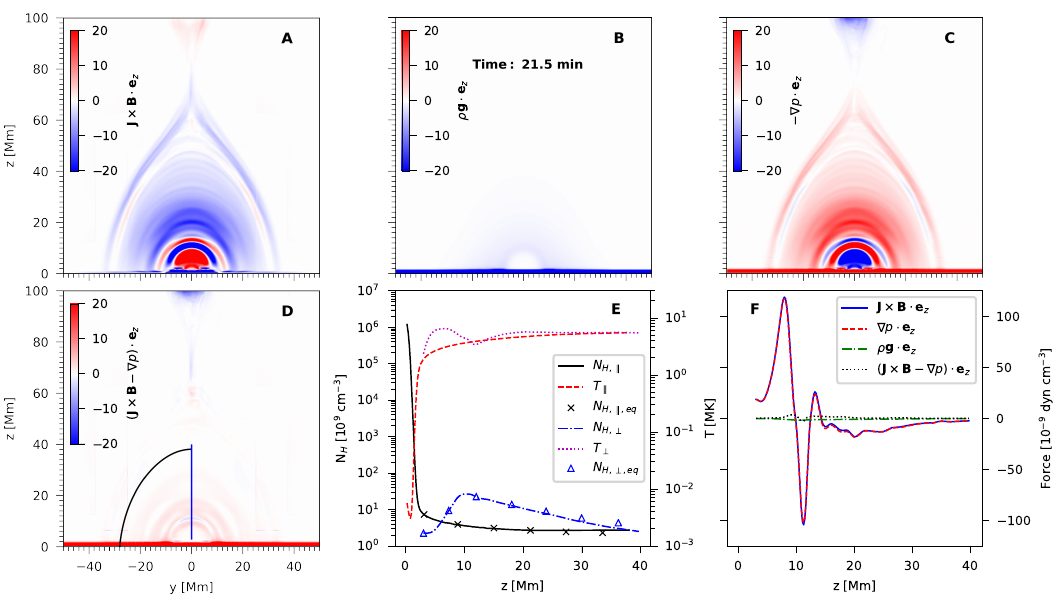}
\caption{ {Force analysis in the flare loop system.} (\textbf{A} to \textbf{C}) The vertical ($z$) components of the ($x$-averaged) Lorentz force, gravity, and pressure gradient force, respectively. The unit of the forces is $10^{-9}\ \mathrm{dyn}\ \mathrm{cm}^{-3}$. (\textbf{D}) The summed vertical components of pressure gradient and Lorentz force. The innermost loops are stressed differently than the outer loops, as the inner loops are low-density loops that formed before the impulsive phase. The red dashed and the black solid line in panel (E) represent the average temperature and density profiles along the black solid magnetic field line in panel (D). Black crosses indicate the expected density distribution derived from the temperature and gravity stratification along that field line. The $x$-averaged temperature and density profiles in the direction perpendicular to the magnetic field lines (i.e. along the vertical blue solid line in panel (D)) are shown in panel (E) by the purple dotted and the blue dashed-dotted line, respectively. The blue triangles represent the expected density distributions derived from the temperature variation and the pressure gradient-Lorentz force balance. The forces along the vertical line in panel (D) are compared in panel (F). An animation of this figure (Movie Fig4) is provided. The animation covers $\sim$7 minutes of physical time starting at t = 14 min (real-time duration 5 s).
\label{Fig3}}
\end{figure}

\begin{figure}
\includegraphics[width=\textwidth]{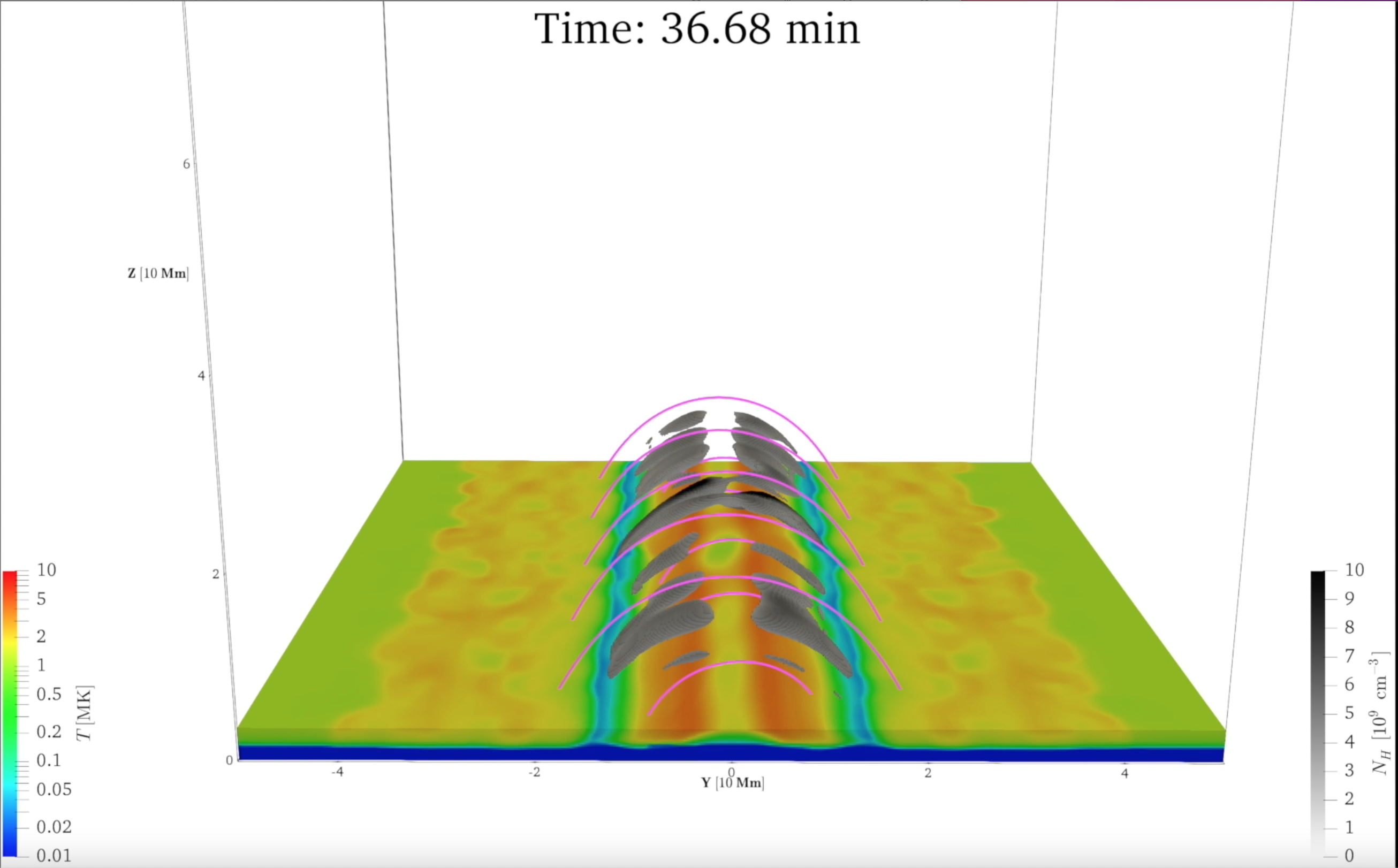}
\caption{{Post-flare coronal rain.} We show regions in darkgrey isosurfaces with temperatures below 25,000 K and of higher than $10^9$ $\mathrm{cm}^{-3}$ proton number density, as well as the lower temperature variation across the transition region and selected post-flare arcade field lines. An animation of this figure (Movie Fig5) is provided, starting at 28.6 minutes, and lasting till 57 minutes. 
\label{FigS7}}
\end{figure}

\begin{figure}
\includegraphics[width=\textwidth]{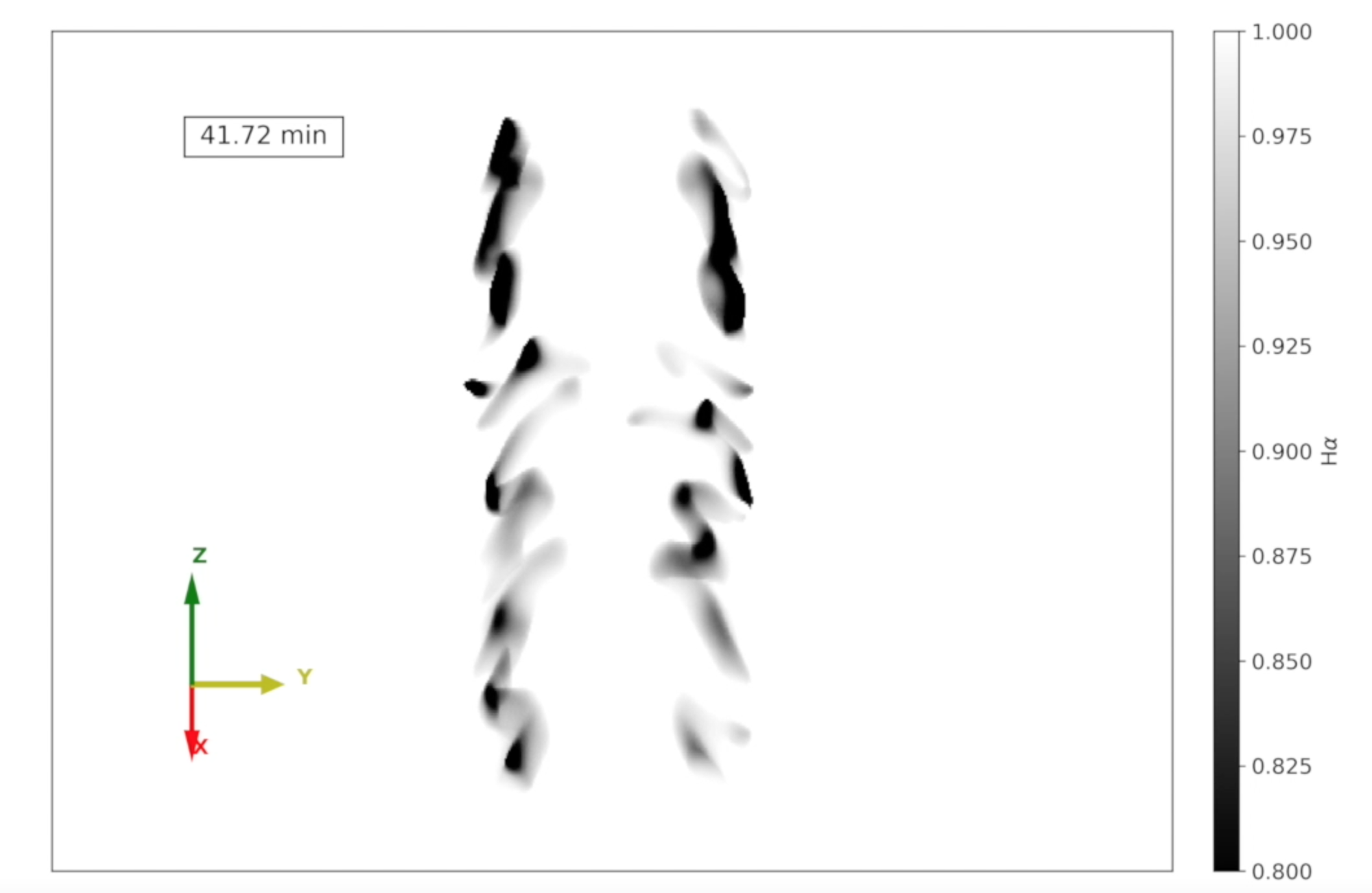}
\caption{{Post-flare coronal rain in an H$\alpha$ proxy.} We show an angled view on the flare loop system.  An animation of this figure (Movie Fig6) is provided, starting at 35.8 minutes, and lasting till 57 minutes. 
\label{FigS8}}
\end{figure}

\begin{figure}
\includegraphics[width=\textwidth]{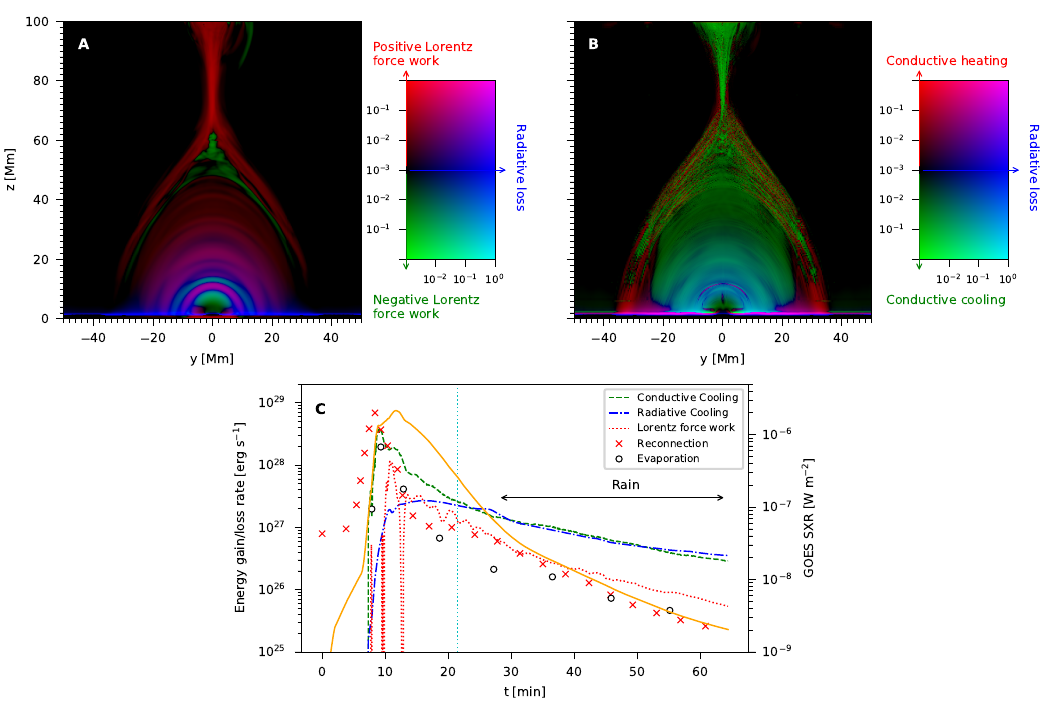}
\caption{Energetic analysis throughout the postflare system. Panel (A) compares spatial distributions of radiative loss and Lorentz force work, while panel (B) compares radiative loss with conductive cooling or heating. For coordinate axis information, refer to Fig.~\ref{Fig3}. The dimensional heating or cooling rate is given in $\mathrm{erg}~\mathrm{ cm}^{-3}~ \mathrm{s}^{-1}$, and uses a multi-color representation that shows the dominant component visually. The time evolution of radiative cooling, conductive cooling, Lorentz force work, energy release rate by reconnection, and the energy flux brought into the flare loop systems by the chromospheric evaporation are shown in panel (C). The cyan vertical dotted line in panel (C) is at the time corresponding to panels (A) and (B). The orange solid curve in panel (C) gives the GOES SXR flux. For detailed computations, we refer to Sect.~\ref{sec:energy}, while an animation (Movie Fig7) is provided. The animation covers $\sim$64 minutes of physical time starting at t = 0 (real-time duration 11 s). 
\label{Fig4}}
\end{figure}

\begin{figure}
\includegraphics[width=\textwidth]{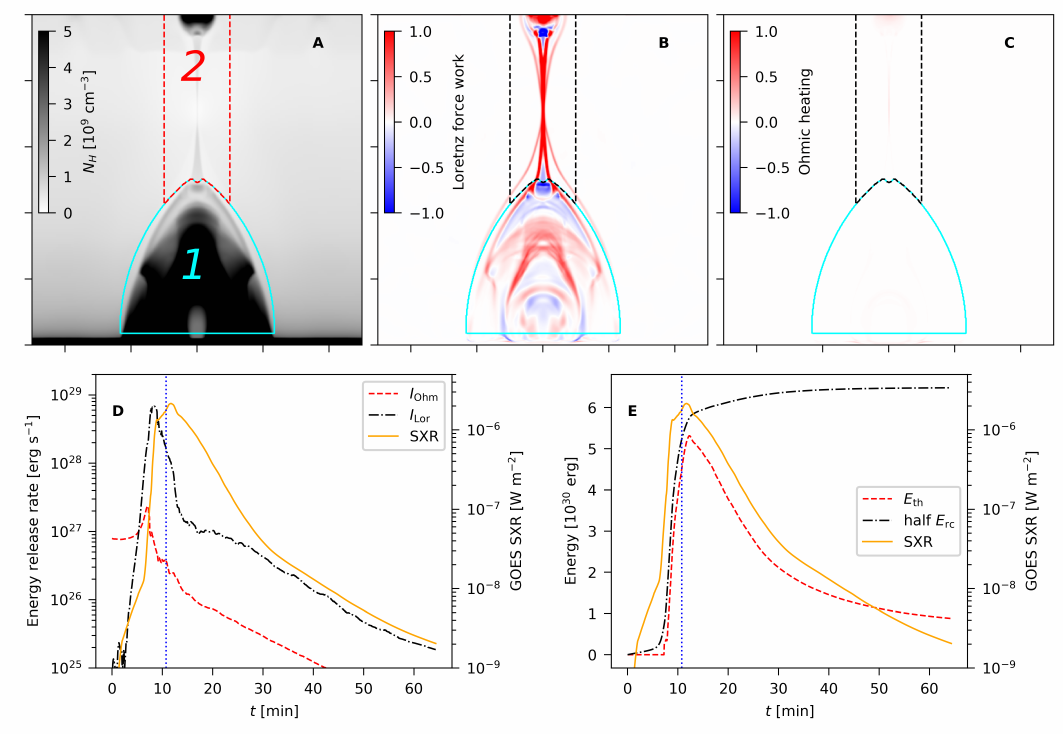}
\caption{ {Energetic analysis of our impulsive flare phase.} Proton number density (A), Lorentz force work (B), and Ohmic heating (C) distributions in the impulsive phase. The area is as shown in Fig.~\ref{FigS1}B.  The time shown in panels (A)-(B)-(C) refers to the vertical dotted line in panels (D) and (E). The time-evolving spatially-integrated (over area 2 in (A)) energy release rates due to Ohmic heating ($I_{\mathrm{Ohm}}$) and Lorentz force work ($I_{\mathrm{Lor}}$) are depicted in panel (D), where the unit is $\mathrm{erg~cm^{-3}~s^{-1}}$. Comparison of the thermal energy ($E_{\mathrm{th}}$) in flare loop systems (over area 1 in (A)) with half of the energy released by magnetic reconnection (half $E_{\mathrm{rc}}$). The GOES SXR flux is given in both panels (D) and (E) by the orange curve.
\label{FigS2}}
\end{figure}

\begin{figure}
\includegraphics[width=\textwidth]{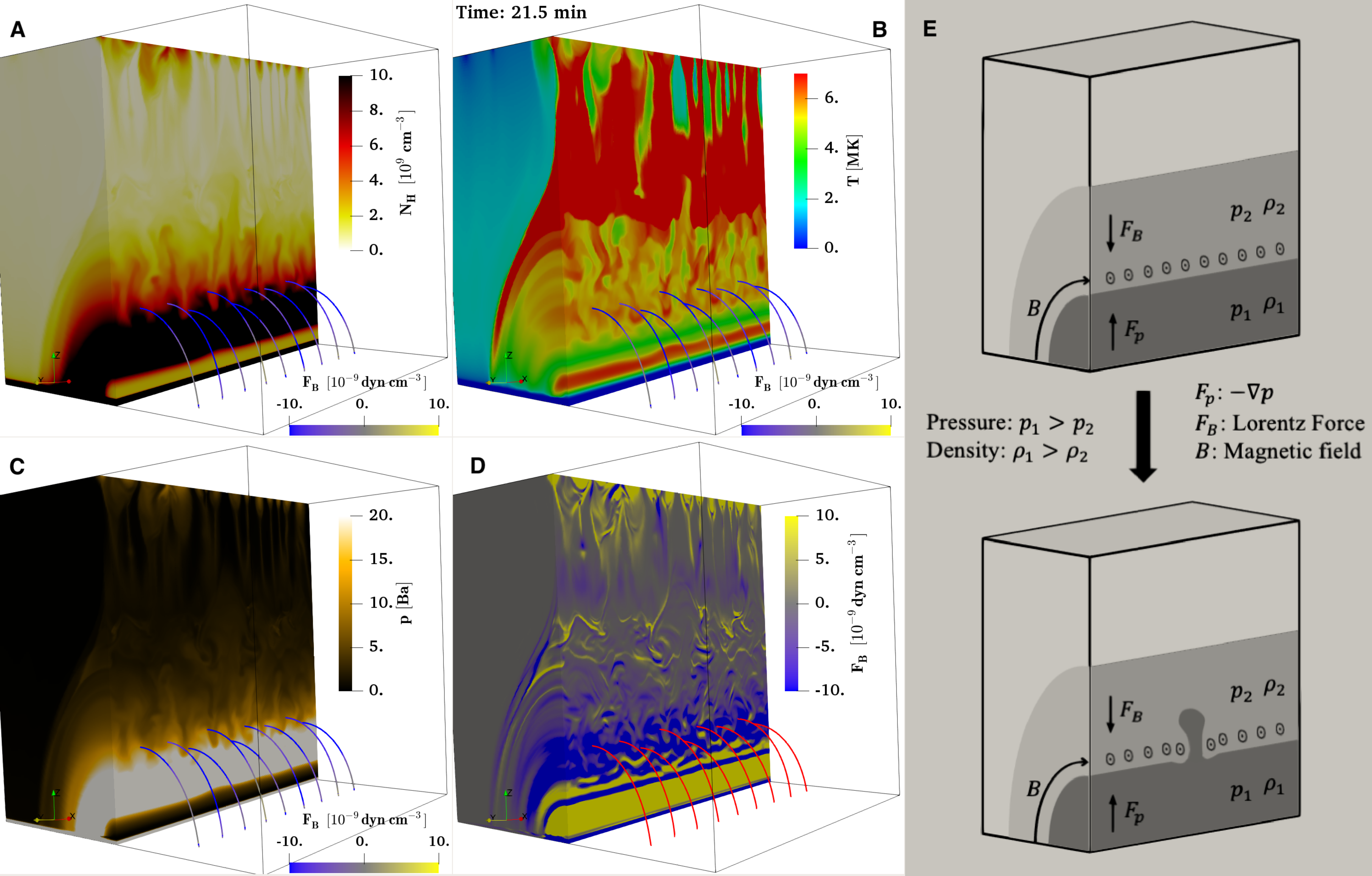}
\caption{ {Rayleigh-Taylor instability inside flare loop systems.} The cartoon in panel (E) illustrates the mechanism. The proton number density, temperature, thermal pressure, and vertical component of the Lorentz force are shown in panels (A) to (D). Magnetic field lines are represented by the curves in panels (A) to (D), and the surface color of the field lines in panels (A) to (C) shows the local vertical component of the Lorentz force. An animation (Movie Fig9) is provided. The animation covers $\sim$14 minutes of physical time starting at t = 11 min (real-time duration 5 s). The proton number density, temperature, thermal pressure, and vertical component of the Lorentz force, EUV 131 {\AA} emissivity, and EUV 193 {\AA} emissivity are shown in panels (A) to (F) of the movie.
\label{Fig5}}
\end{figure}

\begin{figure}
\includegraphics[width=\textwidth]{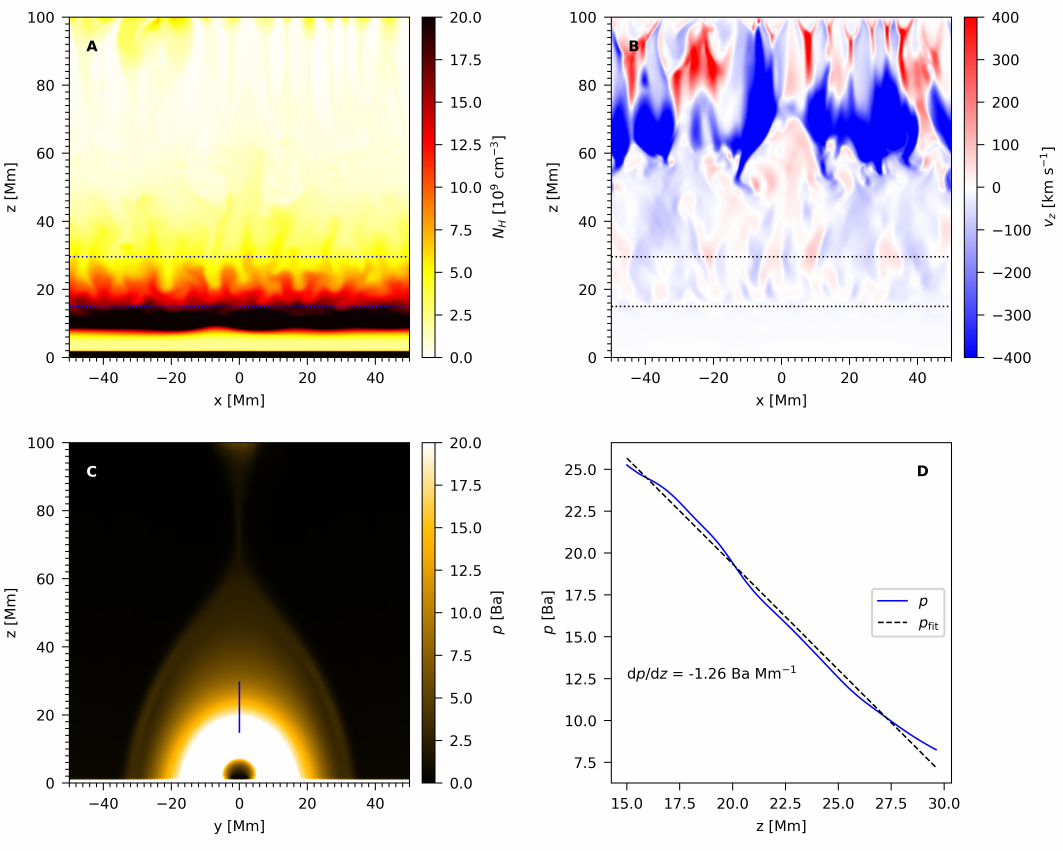}
\caption{ {Instability analysis of gradual phase flare loops.} Proton number density (A) and vertical velocity (B) at slice $y=0$.  Panel (C) shows mean pressure distribution on y-z plane given by Eqn.~(\ref{EqApth}). Panel (D) gives profile of the mean pressure at the vertical line in (C), and the corresponding region in (A) and (B) is the region between the horizontal dotted lines. A fitting of the pressure profile is also given in (D), as well as the pressure gradient from the fitting. The corresponding time is the same as that in Fig.~\ref{Fig5}, that is, $t=21.5\ \mathrm{min}$.
\label{FigS3}}
\end{figure}

In the observation reported in \citet{Jing2016NatSR}, flare-driven coronal rain initially appears in the inner loops, and then the formation locations expands outward over time. Such a process is well captured in our simulation (see the Movie of Fig.~\ref{FigS7}). Our simulation demonstrates that the high density flare loop systems are gradually emptied from the inside out, due to the formation of coronal rain (see the Movie of Fig.~\ref{Fig2}). Possibly due to our limited (order 200 km) simulation resolution, the coronal rain in our simulation does not exhibit many fiber-like fine structures as the observation. For the next step, higher resolution will be employed in the numerical study of the gradual phase to capture more details, and a more realistic magnetic field topology will be employed, such as a magnetic field topology inspired from \citet{Aulanier2012} or a topology extrapolated from a magnetogram. Future work could improve on the neglect of non-thermal plasma aspects that are vital during the impulsive (and still present in the gradual) flaring phase, along the lines of self-consistent models unifying fast electron beam physics with multi-dimensional MHD modeling \citep{Ruan2020ApJ,Druett2023,Druett2024}.

In this article we show the importance of the Lorentz force during the flare evolution, especially in the gradual phase and in the lower-lying postflare loops where ultimately postflare rain developed. For understanding how complex magnetic topologies may be liable to flare onsets, non-linear forcefree field extrapolations remain valuable, and non-linear forcefree field approaches are very useful in modelling coronal magnetic fields \citep{Wiegelmann2012LRSP}. However, the Lorentz force and how it acts to redistribute energy should be important in regions where pressure changes by orders of magnitude on a small length scale, like flare loops. Such kind of regions may only occupy a small part of the corona.
%Future work could improve on the neglect of non-thermal plasma aspects that are vital during the impulsive (and still present in the gradual) flaring phase, along the lines of self-consistent models unifying fast electron beam physics with multi-dimensional MHD modeling \citep{Ruan2020ApJ}.

%% IMPORTANT! The old "\acknowledgment" command has be depreciated. It was
%% not robust enough to handle our new dual anonymous review requirements and
%% thus been replaced with the acknowledgment environment. If you try to 
%% compile with \acknowledgment you will get an error print to the screen
%% and in the compiled pdf.
\begin{acknowledgments}
We thank Dr. Yuhao Zhou for the discussion on coronal rain and the synthesis of the H$\alpha$ images. This project received funding from the European Research Council (ERC) under the European Union’s Horizon 2020 research and innovation program (grant agreement No. 833251 PROMINENT ERC-ADG 2018). This research is supported by Internal funds KU Leuven, project C14/19/089 TRACESpace and FWO project G0B4521N. This research was supported by the International Space Science Institute (ISSI) in Bern, through ISSI International Team project \#545 (“Observe Local Think Global: What Solar Observations can Teach us about Multiphase Plasmas across Physical Scales”). L.Y. is supported by the Youth Innovation Promotion Association of CAS (Grant No. 2021064). P.A. acknowledges funding from his STFC Ernest Rutherford Fellowship (No. ST/R004285/2). Resources and services used in this work were provided by the VSC (Flemish Supercomputer Center), funded by the Research Foundation - Flanders (FWO) and the Flemish Government.
\end{acknowledgments}

%% To help institutions obtain information on the effectiveness of their 
%% telescopes the AAS Journals has created a group of keywords for telescope 
%% facilities.
%
%% Following the acknowledgments section, use the following syntax and the
%% \facility{} or \facilities{} macros to list the keywords of facilities used 
%% in the research for the paper.  Each keyword is check against the master 
%% list during copy editing.  Individual instruments can be provided in 
%% parentheses, after the keyword, but they are not verified.

\vspace{5mm}
\facilities{SDO(AIA)}

%% Similar to \facility{}, there is the optional \software command to allow 
%% authors a place to specify which programs were used during the creation of 
%% the manuscript. Authors should list each code and include either a
%% citation or url to the code inside ()s when available.

\software{MPI-AMRVAC \citep{Keppens2023}          }

%% Appendix material should be preceded with a single \appendix command.
%% There should be a \section command for each appendix. Mark appendix
%% subsections with the same markup you use in the main body of the paper.

%% Each Appendix (indicated with \section) will be lettered A, B, C, etc.
%% The equation counter will reset when it encounters the \appendix
%% command and will number appendix equations (A1), (A2), etc. The
%% Figure and Table counter will not reset.

%\caption{ {Energetic analysis throughout the postflare system.} Panel (A) compares spatial distributions of radiative loss and Lorentz force work, while panel (B) compares radiative loss with conductive cooling or heating. For coordinate axis information, refer to Fig.~\ref{Fig3}. The dimensional heating or cooling rate is given in $\mathrm{erg}~\mathrm{ cm}^{-3}~ \mathrm{s}^{-1}$, and uses a multi-color representation that shows the dominant component visually. The time evolution of radiative cooling, conductive cooling, Lorentz force work, energy release rate by reconnection, and the energy flux brought into the flare loop systems by the chromospheric evaporation are shown in panel (C). The cyan vertical dotted line in panel (C) is at the time corresponding to panels (A) and (B). The orange solid curve in panel (C) gives the GOES SXR flux. For detailed computations, we refer to Sect.~\ref{sec:energy}, while an animation (Movie Fig7) is provided. The animation covers $\sim$64 minutes of physical time starting at t = 0 (real-time duration 11 s). Cold rain material (T $<$ 50, 000 K) distribution is shown in panel (B) of the movie with white contours at column depths $10^{18}$, $10^{19}$ cm$^{−2}$.}

\clearpage

\appendix

\section{Mean Lorentz force calculated from original current and magnetic field}\label{sec:meanforce}

\begin{figure}
\includegraphics[width=\textwidth]{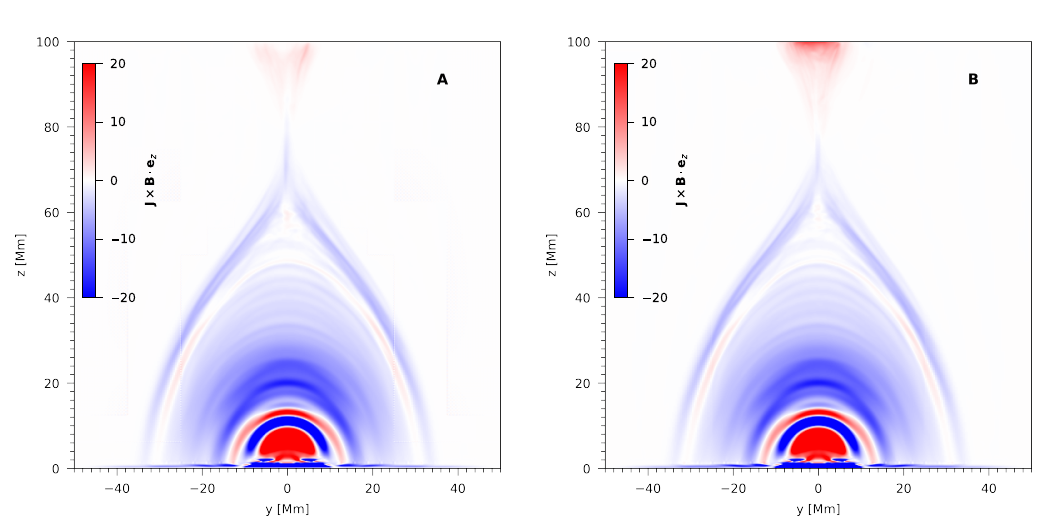}
\caption{ Mean Lorentz force given by two different methods. (\textbf{A}) Mean Lorentz force calculated according to the method in Sect.~\ref{sec:force}. (\textbf{B}) Mean Lorentz force calculated according to the method in Appendix~\ref{sec:meanforce}. The unit of the force is $10^{-9}\ \mathrm{dyn}\ \mathrm{cm}^{-3}$ and the corresponding time is $t=21.5~\mathrm{min}$.
\label{Fig11}}
\end{figure}

The z-component of Lorentz force is given by
\begin{equation}
F_{B,z} (x,y,z) = J_x (x,y,z) B_y (x,y,z) - J_y (x,y,z) B_x (x,y,z),
\end{equation}
where the current is given by
\begin{equation}
\mathbf{J} (x,y,x) = \nabla \times \mathbf{B} (x,y,z) .
\end{equation}
The mean Lorentz force in the z direction is given by
\begin{equation}
\bar{F}_{B,z} (y,z) = \int_{x_{\mathrm{min}}}^{x_{\mathrm{max}}} F_{B,z} (x,y,z) \mathrm{d}x / (x_{\mathrm{max}} - x_{\mathrm{min}})\,. 
\end{equation}
Panel (B) of Fig.~\ref{Fig11} shows the mean Lorentz force given by this method, where the mean force given by the method in section \ref{sec:force} is demonstrated in panel (A) for comparison. The difference between the results given by the two methods is minor.

\bibliography{reference}{}

\begin{thebibliography}{}
\expandafter\ifx\csname natexlab\endcsname\relax\def\natexlab#1{#1}\fi
\providecommand{\url}[1]{\href{#1}{#1}}
\providecommand{\dodoi}[1]{doi:~\href{http://doi.org/#1}{\nolinkurl{#1}}}
\providecommand{\doeprint}[1]{\href{http://ascl.net/#1}{\nolinkurl{http://ascl.net/#1}}}
\providecommand{\doarXiv}[1]{\href{https://arxiv.org/abs/#1}{\nolinkurl{https://arxiv.org/abs/#1}}}

\bibitem[{{Aliu} {et~al.}(2012){Aliu}, {Arlen}, {Aune}, {Beilicke}, {Benbow},
  {Bouvier}, {Bradbury}, {Buckley}, {Bugaev}, {Byrum}, {Cannon}, {Cesarini},
  {Ciupik}, {Collins-Hughes}, {Connolly}, {Cui}, {Dickherber}, {Duke},
  {Errando}, {Falcone}, {Finley}, {Finnegan}, {Fortson}, {Furniss}, {Galante},
  {Gall}, {Godambe}, {Griffin}, {Grube}, {Guenette}, {Gyuk}, {Hanna}, {Holder},
  {Huan}, {Hughes}, {Hui}, {Humensky}, {Imran}, {Kaaret}, {Karlsson},
  {Kertzman}, {Kieda}, {Krawczynski}, {Krennrich}, {Lang}, {LeBohec},
  {Madhavan}, {Maier}, {Majumdar}, {McArthur}, {McCann}, {Moriarty},
  {Mukherjee}, {Nu{\~n}ez}, {Ong}, {Orr}, {Otte}, {Park}, {Perkins}, {Pichel},
  {Pohl}, {Prokoph}, {Quinn}, {Ragan}, {Reyes}, {Reynolds}, {Roache}, {Rose},
  {Ruppel}, {Saxon}, {Schroedter}, {Sembroski}, {{\c{S}}ent{\"u}rk}, {Skole},
  {Staszak}, {Te{\v{s}}i{\'c}}, {Theiling}, {Thibadeau}, {Tsurusaki}, {Tyler},
  {Varlotta}, {Vassiliev}, {Vincent}, {Vivier}, {Wakely}, {Ward}, {Weekes},
  {Weinstein}, {Weisgarber}, {Williams}, \& {Zitzer}}]{Aliu2012ApJ}
{Aliu}, E., {Arlen}, T., {Aune}, T., {et~al.} 2012, The Astrophys. J., 746,
  141, \dodoi{10.1088/0004-637X/746/2/141}

\bibitem[{{Aschwanden} {et~al.}(2017){Aschwanden}, {Caspi}, {Cohen}, {Holman},
  {Jing}, {Kretzschmar}, {Kontar}, {McTiernan}, {Mewaldt}, {O'Flannagain},
  {Richardson}, {Ryan}, {Warren}, \& {Xu}}]{Aschwanden2017ApJ}
{Aschwanden}, M.~J., {Caspi}, A., {Cohen}, C. M.~S., {et~al.} 2017, The
  Astrophys. J., 836, 17, \dodoi{10.3847/1538-4357/836/1/17}

\bibitem[{{Aulanier} {et~al.}(2012){Aulanier}, {Janvier}, \&
  {Schmieder}}]{Aulanier2012}
{Aulanier}, G., {Janvier}, M., \& {Schmieder}, B. 2012, \aap, 543, A110,
  \dodoi{10.1051/0004-6361/201219311}

\bibitem[{{Avrett} \& {Loeser}(2008)}]{Avrett2008ApJS}
{Avrett}, E.~H., \& {Loeser}, R. 2008, Astrophys. J. Suppl. Ser., 175, 229,
  \dodoi{10.1086/523671}

\bibitem[{{Benz}(2017)}]{Benz2017LRSP}
{Benz}, A.~O. 2017, Living Rev. Sol. Phys., 14, 2,
  \dodoi{10.1007/s41116-016-0004-3}

\bibitem[{{Bian} {et~al.}(2016{\natexlab{a}}){Bian}, {Kontar}, \&
  {Emslie}}]{Bian2016ApJa}
{Bian}, N.~H., {Kontar}, E.~P., \& {Emslie}, A.~G. 2016{\natexlab{a}}, \apj,
  824, 78, \dodoi{10.3847/0004-637X/824/2/78}

\bibitem[{{Bian} {et~al.}(2016{\natexlab{b}}){Bian}, {Watters}, {Kontar}, \&
  {Emslie}}]{Bian2016ApJb}
{Bian}, N.~H., {Watters}, J.~M., {Kontar}, E.~P., \& {Emslie}, A.~G.
  2016{\natexlab{b}}, \apj, 833, 76, \dodoi{10.3847/1538-4357/833/1/76}

\bibitem[{{Birn} {et~al.}(2009){Birn}, {Fletcher}, {Hesse}, \&
  {Neukirch}}]{Birn2009ApJ}
{Birn}, J., {Fletcher}, L., {Hesse}, M., \& {Neukirch}, T. 2009, The Astrophys.
  J., 695, 1151, \dodoi{10.1088/0004-637X/695/2/1151}

\bibitem[{{Bradshaw} \& {Cargill}(2005)}]{Bradshaw2005A&A}
{Bradshaw}, S.~J., \& {Cargill}, P.~J. 2005, Astron. \& Astrophys., 437, 311,
  \dodoi{10.1051/0004-6361:20042405}

\bibitem[{{Bradshaw} \& {Cargill}(2010)}]{Bradshaw2010ApJ}
---. 2010, The Astrophys. J., 717, 163, \dodoi{10.1088/0004-637X/717/1/163}

\bibitem[{{Brooks} {et~al.}(2024){Brooks}, {Reep}, {Ugarte-Urra}, {Unverferth},
  \& {Warren}}]{Brooks2024}
{Brooks}, D.~H., {Reep}, J.~W., {Ugarte-Urra}, I., {Unverferth}, J.~E., \&
  {Warren}, H.~P. 2024, \apj, 962, 105, \dodoi{10.3847/1538-4357/ad18be}

\bibitem[{{Cargill} \& {Bradshaw}(2013)}]{Cargill2013ApJ}
{Cargill}, P.~J., \& {Bradshaw}, S.~J. 2013, The Astrophys. J., 772, 40,
  \dodoi{10.1088/0004-637X/772/1/40}

\bibitem[{{Cargill} \& {Priest}(1983)}]{Cargill1983ApJ}
{Cargill}, P.~J., \& {Priest}, E.~R. 1983, The Astrophys. J., 266, 383,
  \dodoi{10.1086/160786}

\bibitem[{{Colgan} {et~al.}(2008){Colgan}, {Abdallah}, {Sherrill}, {Foster},
  {Fontes}, \& {Feldman}}]{Colgan2008ApJ}
{Colgan}, J., {Abdallah}, J., J., {Sherrill}, M.~E., {et~al.} 2008, The
  Astrophys. J., 689, 585, \dodoi{10.1086/592561}

\bibitem[{{Del Zanna} {et~al.}(2015){Del Zanna}, {Dere}, {Young}, {Landi}, \&
  {Mason}}]{DelZanna2015A&A}
{Del Zanna}, G., {Dere}, K.~P., {Young}, P.~R., {Landi}, E., \& {Mason}, H.~E.
  2015, Astron. \& Astrophys., 582, A56, \dodoi{10.1051/0004-6361/201526827}

\bibitem[{{Druett} {et~al.}(2023{\natexlab{a}}){Druett}, {Ruan}, \&
  {Keppens}}]{Druett2024}
{Druett}, M., {Ruan}, W., \& {Keppens}, R. 2023{\natexlab{a}}, arXiv e-prints,
  arXiv:2310.09939, \dodoi{10.48550/arXiv.2310.09939}

\bibitem[{{Druett} {et~al.}(2023{\natexlab{b}}){Druett}, {Ruan}, \&
  {Keppens}}]{Druett2023}
{Druett}, M.~K., {Ruan}, W., \& {Keppens}, R. 2023{\natexlab{b}}, \solphys,
  298, 134, \dodoi{10.1007/s11207-023-02224-4}

\bibitem[{{Emslie} \& {Bian}(2018)}]{Emslie2018ApJ}
{Emslie}, A.~G., \& {Bian}, N.~H. 2018, \apj, 865, 67,
  \dodoi{10.3847/1538-4357/aad961}

\bibitem[{{Field}(1965)}]{Field1965ApJ}
{Field}, G.~B. 1965, The Astrophys. J., 142, 531, \dodoi{10.1086/148317}

\bibitem[{{Fleishman} {et~al.}(2020){Fleishman}, {Gary}, {Chen}, {Kuroda},
  {Yu}, \& {Nita}}]{Fleishman2020Sci}
{Fleishman}, G.~D., {Gary}, D.~E., {Chen}, B., {et~al.} 2020, Science, 367,
  278, \dodoi{10.1126/science.aax6874}

\bibitem[{{Fletcher} {et~al.}(2011){Fletcher}, {Dennis}, {Hudson}, {Krucker},
  {Phillips}, {Veronig}, {Battaglia}, {Bone}, {Caspi}, {Chen}, {Gallagher},
  {Grigis}, {Ji}, {Liu}, {Milligan}, \& {Temmer}}]{Fletcher2011SSRv}
{Fletcher}, L., {Dennis}, B.~R., {Hudson}, H.~S., {et~al.} 2011, Space Sci.
  Rev., 159, 19, \dodoi{10.1007/s11214-010-9701-8}

\bibitem[{{Freeland} \& {Handy}(1998)}]{Freeland1998SoPh}
{Freeland}, S.~L., \& {Handy}, B.~N. 1998, Sol. Phys., 182, 497,
  \dodoi{10.1023/A:1005038224881}

\bibitem[{{G{\"u}nther} {et~al.}(2020){G{\"u}nther}, {Zhan}, {Seager},
  {Rimmer}, {Ranjan}, {Stassun}, {Oelkers}, {Daylan}, {Newton}, {Kristiansen},
  {Olah}, {Gillen}, {Rappaport}, {Ricker}, {Vanderspek}, {Latham}, {Winn},
  {Jenkins}, {Glidden}, {Fausnaugh}, {Levine}, {Dittmann}, {Quinn},
  {Krishnamurthy}, \& {Ting}}]{Gunther2020AJ}
{G{\"u}nther}, M.~N., {Zhan}, Z., {Seager}, S., {et~al.} 2020, Astron. J., 159,
  60, \dodoi{10.3847/1538-3881/ab5d3a}

\bibitem[{{Guo} {et~al.}(2014){Guo}, {Huang}, {Bhattacharjee}, \&
  {Innes}}]{Guo2014ApJ}
{Guo}, L.~J., {Huang}, Y.~M., {Bhattacharjee}, A., \& {Innes}, D.~E. 2014, The
  Astrophys. J. Lett., 796, L29, \dodoi{10.1088/2041-8205/796/2/L29}

\bibitem[{Harten {et~al.}(1983)Harten, Lax, \& Leer}]{Harten1983upstream}
Harten, A., Lax, P.~D., \& Leer, B.~v. 1983, SIAM review, 25, 35

\bibitem[{{Heinzel} {et~al.}(2015){Heinzel}, {Gun{\'a}r}, \&
  {Anzer}}]{Heinzel2015A&A}
{Heinzel}, P., {Gun{\'a}r}, S., \& {Anzer}, U. 2015, Astron. \& Astrophys.,
  579, A16, \dodoi{10.1051/0004-6361/201525716}

\bibitem[{{Hermans} \& {Keppens}(2021)}]{Hermans2021A&A}
{Hermans}, J., \& {Keppens}, R. 2021, Astron. \& Astrophys., 655, A36,
  \dodoi{10.1051/0004-6361/202140665}

\bibitem[{{Hudson}(2011)}]{Hudson2011SSRv}
{Hudson}, H.~S. 2011, Space Sci. Rev., 158, 5,
  \dodoi{10.1007/s11214-010-9721-4}

\bibitem[{{Jing} {et~al.}(2016){Jing}, {Xu}, {Cao}, {Liu}, {Gary}, \&
  {Wang}}]{Jing2016NatSR}
{Jing}, J., {Xu}, Y., {Cao}, W., {et~al.} 2016, Scientific Reports, 6, 24319,
  \dodoi{10.1038/srep24319}

\bibitem[{{Johnston} {et~al.}(2020){Johnston}, {Cargill}, {Hood}, {De Moortel},
  {Bradshaw}, \& {Vaseekar}}]{Johnston2020A&A}
{Johnston}, C.~D., {Cargill}, P.~J., {Hood}, A.~W., {et~al.} 2020, Astron. \&
  Astrophys., 635, A168, \dodoi{10.1051/0004-6361/201936979}

\bibitem[{{Keppens} {et~al.}(2023){Keppens}, {Popescu Braileanu}, {Zhou},
  {Ruan}, {Xia}, {Guo}, {Claes}, \& {Bacchini}}]{Keppens2023}
{Keppens}, R., {Popescu Braileanu}, B., {Zhou}, Y., {et~al.} 2023, Astron. \&
  Astrophys., 673, A66, \dodoi{10.1051/0004-6361/202245359}

\bibitem[{{Khan} {et~al.}(2007){Khan}, {Bain}, \& {Fletcher}}]{Khan2007A&A}
{Khan}, J.~I., {Bain}, H.~M., \& {Fletcher}, L. 2007, Astron. \& Astrophys.,
  475, 333, \dodoi{10.1051/0004-6361:20077894}

\bibitem[{{Lemen} {et~al.}(2012){Lemen}, {Title}, {Akin}, {Boerner}, {Chou},
  {Drake}, {Duncan}, {Edwards}, {Friedlaender}, {Heyman}, {Hurlburt}, {Katz},
  {Kushner}, {Levay}, {Lindgren}, {Mathur}, {McFeaters}, {Mitchell}, {Rehse},
  {Schrijver}, {Springer}, {Stern}, {Tarbell}, {Wuelser}, {Wolfson}, {Yanari},
  {Bookbinder}, {Cheimets}, {Caldwell}, {Deluca}, {Gates}, {Golub}, {Park},
  {Podgorski}, {Bush}, {Scherrer}, {Gummin}, {Smith}, {Auker}, {Jerram},
  {Pool}, {Soufli}, {Windt}, {Beardsley}, {Clapp}, {Lang}, \&
  {Waltham}}]{Lemen2012SoPh}
{Lemen}, J.~R., {Title}, A.~M., {Akin}, D.~J., {et~al.} 2012, Sol. Phys., 275,
  17, \dodoi{10.1007/s11207-011-9776-8}

\bibitem[{{Li} {et~al.}(2014){Li}, {Ding}, {Guo}, \& {Dai}}]{Li2014ApJ}
{Li}, Y., {Ding}, M.~D., {Guo}, Y., \& {Dai}, Y. 2014, The Astrophys. J., 793,
  85, \dodoi{10.1088/0004-637X/793/2/85}

\bibitem[{{Liu} {et~al.}(2013){Liu}, {Zhang}, {Wang}, \& {Cheng}}]{Liu2013ApJ}
{Liu}, K., {Zhang}, J., {Wang}, Y., \& {Cheng}, X. 2013, The Astrophys. J.,
  768, 150, \dodoi{10.1088/0004-637X/768/2/150}

\bibitem[{{Mart{\'\i}nez Oliveros} {et~al.}(2022){Mart{\'\i}nez Oliveros},
  {Guevara G{\'o}mez}, {Saint-Hilaire}, {Hudson}, \&
  {Krucker}}]{Martinez2022ApJ}
{Mart{\'\i}nez Oliveros}, J.~C., {Guevara G{\'o}mez}, J.~C., {Saint-Hilaire},
  P., {Hudson}, H., \& {Krucker}, S. 2022, The Astrophys. J., 936, 56,
  \dodoi{10.3847/1538-4357/ac83b7}

\bibitem[{{Mason} \& {Kniezewski}(2022)}]{Mason2022ApJ}
{Mason}, E.~I., \& {Kniezewski}, K.~L. 2022, The Astrophys. J., 939, 21,
  \dodoi{10.3847/1538-4357/ac94d7}

\bibitem[{{McKenzie} \& {Hudson}(1999)}]{McKenzie1999ApJ}
{McKenzie}, D.~E., \& {Hudson}, H.~S. 1999, The Astrophys. J. Lett., 519, L93,
  \dodoi{10.1086/312110}

\bibitem[{{McKenzie} \& {Savage}(2009)}]{McKenzie2009ApJ}
{McKenzie}, D.~E., \& {Savage}, S.~L. 2009, The Astrophys. J., 697, 1569,
  \dodoi{10.1088/0004-637X/697/2/1569}

\bibitem[{{Pinto} {et~al.}(2015){Pinto}, {Vilmer}, \& {Brun}}]{Pinto2015A&A}
{Pinto}, R.~F., {Vilmer}, N., \& {Brun}, A.~S. 2015, Astron. \& Astrophys.,
  576, A37, \dodoi{10.1051/0004-6361/201323358}

\bibitem[{{Pontin} \& {Priest}(2022)}]{Pontin2022LRSP}
{Pontin}, D.~I., \& {Priest}, E.~R. 2022, Living Rev. Sol. Phys., 19, 1,
  \dodoi{10.1007/s41116-022-00032-9}

\bibitem[{{Priest} \& {Forbes}(2002)}]{Priest2002A&ARv}
{Priest}, E.~R., \& {Forbes}, T.~G. 2002, The Astron. Astrophys. Rev., 10, 313,
  \dodoi{10.1007/s001590100013}

\bibitem[{{Reep} {et~al.}(2020){Reep}, {Antolin}, \& {Bradshaw}}]{Reep2020ApJ}
{Reep}, J.~W., {Antolin}, P., \& {Bradshaw}, S.~J. 2020, The Astrophys. J.,
  890, 100, \dodoi{10.3847/1538-4357/ab6bdc}

\bibitem[{{Reep} \& {Knizhnik}(2019)}]{Reep2019ApJ}
{Reep}, J.~W., \& {Knizhnik}, K.~J. 2019, The Astrophys. J., 874, 157,
  \dodoi{10.3847/1538-4357/ab0ae7}

\bibitem[{{Reep} \& {Toriumi}(2017)}]{Reep2017ApJ}
{Reep}, J.~W., \& {Toriumi}, S. 2017, The Astrophys. J., 851, 4,
  \dodoi{10.3847/1538-4357/aa96fe}

\bibitem[{{Reep} {et~al.}(2022){Reep}, {Ugarte-Urra}, {Warren}, \&
  {Barnes}}]{Reep2022ApJ}
{Reep}, J.~W., {Ugarte-Urra}, I., {Warren}, H.~P., \& {Barnes}, W.~T. 2022, The
  Astrophys. J., 933, 106, \dodoi{10.3847/1538-4357/ac7398}

\bibitem[{{Reeves} {et~al.}(2017){Reeves}, {Freed}, {McKenzie}, \&
  {Savage}}]{Reeves2017ApJ}
{Reeves}, K.~K., {Freed}, M.~S., {McKenzie}, D.~E., \& {Savage}, S.~L. 2017,
  The Astrophys. J., 836, 55, \dodoi{10.3847/1538-4357/836/1/55}

\bibitem[{{Reeves} {et~al.}(2010){Reeves}, {Linker}, {Miki{\'c}}, \&
  {Forbes}}]{Reeves2010ApJ}
{Reeves}, K.~K., {Linker}, J.~A., {Miki{\'c}}, Z., \& {Forbes}, T.~G. 2010,
  \apj, 721, 1547, \dodoi{10.1088/0004-637X/721/2/1547}

\bibitem[{{Rempel} {et~al.}(2023){Rempel}, {Chintzoglou}, {Cheung}, {Fan}, \&
  {Kleint}}]{Rempel2023}
{Rempel}, M., {Chintzoglou}, G., {Cheung}, M. C.~M., {Fan}, Y., \& {Kleint}, L.
  2023, \apj, 955, 105, \dodoi{10.3847/1538-4357/aced4d}

\bibitem[{{Ruan} {et~al.}(2020){Ruan}, {Xia}, \& {Keppens}}]{Ruan2020ApJ}
{Ruan}, W., {Xia}, C., \& {Keppens}, R. 2020, The Astrophys. J., 896, 97,
  \dodoi{10.3847/1538-4357/ab93db}

\bibitem[{{Ruan} {et~al.}(2023){Ruan}, {Yan}, \& {Keppens}}]{Ruan2023ApJ}
{Ruan}, W., {Yan}, L., \& {Keppens}, R. 2023, The Astrophys. J., 947, 67,
  \dodoi{10.3847/1538-4357/ac9b4e}

\bibitem[{{Ruan} {et~al.}(2021){Ruan}, {Zhou}, \& {Keppens}}]{Ruan2021ApJ}
{Ruan}, W., {Zhou}, Y., \& {Keppens}, R. 2021, The Astrophys. J. Lett., 920,
  L15, \dodoi{10.3847/2041-8213/ac27b0}

\bibitem[{{Samanta} {et~al.}(2021){Samanta}, {Tian}, {Chen}, {Reeves},
  {Cheung}, {Vourlidas}, \& {Banerjee}}]{Samanta2021Innov}
{Samanta}, T., {Tian}, H., {Chen}, B., {et~al.} 2021, The Innovation, 2,
  100083, \dodoi{10.1016/j.xinn.2021.100083}

\bibitem[{{Savage} \& {McKenzie}(2011)}]{Savage2011ApJ}
{Savage}, S.~L., \& {McKenzie}, D.~E. 2011, \apj, 730, 98,
  \dodoi{10.1088/0004-637X/730/2/98}

\bibitem[{{Schmieder} {et~al.}(1995){Schmieder}, {Heinzel}, {Wiik}, {Lemen},
  {Anwar}, {Kotrc}, \& {Hiei}}]{Schmieder1995SoPh}
{Schmieder}, B., {Heinzel}, P., {Wiik}, J.~E., {et~al.} 1995, Sol. Phys., 156,
  337, \dodoi{10.1007/BF00670231}

\bibitem[{{Shen} {et~al.}(2022){Shen}, {Chen}, {Reeves}, {Yu}, {Polito}, \&
  {Xie}}]{Shen2022NatAs}
{Shen}, C., {Chen}, B., {Reeves}, K.~K., {et~al.} 2022, Nat. Astron., 6, 317,
  \dodoi{10.1038/s41550-021-01570-2}

\bibitem[{{Shibata} \& {Magara}(2011)}]{Shibata2011LRSP}
{Shibata}, K., \& {Magara}, T. 2011, Living Rev. Sol. Phys., 8, 6,
  \dodoi{10.12942/lrsp-2011-6}

\bibitem[{{Tan} {et~al.}(2023){Tan}, {Hou}, \& {Tian}}]{Tan2023MNRAS}
{Tan}, G., {Hou}, Y., \& {Tian}, H. 2023, Mon. Notices Royal Astron. Soc.,
  \dodoi{10.1093/mnras/stad1228}

\bibitem[{{Toriumi} {et~al.}(2017){Toriumi}, {Schrijver}, {Harra}, {Hudson}, \&
  {Nagashima}}]{Toriumi2017ApJ}
{Toriumi}, S., {Schrijver}, C.~J., {Harra}, L.~K., {Hudson}, H., \&
  {Nagashima}, K. 2017, The Astrophys. J., 834, 56,
  \dodoi{10.3847/1538-4357/834/1/56}

\bibitem[{{Tsurutani} {et~al.}(2009){Tsurutani}, {Verkhoglyadova}, {Mannucci},
  {Lakhina}, {Li}, \& {Zank}}]{Tsurutani2009RaSc}
{Tsurutani}, B.~T., {Verkhoglyadova}, O.~P., {Mannucci}, A.~J., {et~al.} 2009,
  Radio Science, 44, RS0A17, \dodoi{10.1029/2008RS004029}

\bibitem[{{van Leer}(1974)}]{vanLeer1974JCoPh}
{van Leer}, B. 1974, J. Comput. Phys., 14, 361,
  \dodoi{10.1016/0021-9991(74)90019-9}

\bibitem[{{{\v{C}}ada} \& {Torrilhon}(2009)}]{Cada2009JCoPh}
{{\v{C}}ada}, M., \& {Torrilhon}, M. 2009, J. Comput. Phys., 228, 4118,
  \dodoi{10.1016/j.jcp.2009.02.020}

\bibitem[{{Wang} {et~al.}(2023){Wang}, {Cheng}, {Ding}, {Liu}, {Liu}, \&
  {Zhu}}]{Wang2023}
{Wang}, Y., {Cheng}, X., {Ding}, M., {et~al.} 2023, arXiv e-prints,
  arXiv:2308.10494, \dodoi{10.48550/arXiv.2308.10494}

\bibitem[{{Warren} {et~al.}(2011){Warren}, {O'Brien}, \&
  {Sheeley}}]{Warren2011ApJ}
{Warren}, H.~P., {O'Brien}, C.~M., \& {Sheeley}, Neil~R., J. 2011, \apj, 742,
  92, \dodoi{10.1088/0004-637X/742/2/92}

\bibitem[{{Wiegelmann} \& {Sakurai}(2012)}]{Wiegelmann2012LRSP}
{Wiegelmann}, T., \& {Sakurai}, T. 2012, Living Reviews in Solar Physics, 9, 5,
  \dodoi{10.12942/lrsp-2012-5}

\bibitem[{{Xia} {et~al.}(2018){Xia}, {Teunissen}, {El Mellah}, {Chan{\'e}}, \&
  {Keppens}}]{Xia2018ApJS}
{Xia}, C., {Teunissen}, J., {El Mellah}, I., {Chan{\'e}}, E., \& {Keppens}, R.
  2018, Astrophys. J. Suppl. Ser., 234, 30, \dodoi{10.3847/1538-4365/aaa6c8}

\bibitem[{{Xie} \& {Reeves}(2023)}]{Xie2023ApJ}
{Xie}, X., \& {Reeves}, K.~K. 2023, \apj, 942, 28,
  \dodoi{10.3847/1538-4357/ac9f47}

\bibitem[{{Xie} {et~al.}(2022){Xie}, {Reeves}, {Shen}, \&
  {Ingram}}]{Xie2022ApJ}
{Xie}, X., {Reeves}, K.~K., {Shen}, C., \& {Ingram}, J.~D. 2022, \apj, 933, 15,
  \dodoi{10.3847/1538-4357/ac695d}

\bibitem[{{Ye} {et~al.}(2023){Ye}, {Raymond}, {Mei}, {Cai}, {Chen}, {Li}, \&
  {Lin}}]{Ye2023}
{Ye}, J., {Raymond}, J.~C., {Mei}, Z., {et~al.} 2023, arXiv e-prints,
  arXiv:2308.09496, \dodoi{10.48550/arXiv.2308.09496}

\bibitem[{{Yokoyama} \& {Shibata}(2001)}]{Yokoyama2001ApJ}
{Yokoyama}, T., \& {Shibata}, K. 2001, The Astrophys. J., 549, 1160,
  \dodoi{10.1086/319440}

\bibitem[{{Zhou} {et~al.}(2021){Zhou}, {Ruan}, {Xia}, \&
  {Keppens}}]{Zhou2021A&A}
{Zhou}, Y.-H., {Ruan}, W.-Z., {Xia}, C., \& {Keppens}, R. 2021, Astron. \&
  Astrophys., 648, A29, \dodoi{10.1051/0004-6361/202040254}

\end{thebibliography}
\bibliographystyle{aasjournal}

%% This command is needed to show the entire author+affiliation list when
%% the collaboration and author truncation commands are used.  It has to
%% go at the end of the manuscript.
%\allauthors

%% Include this line if you are using the \added, \replaced, \deleted
%% commands to see a summary list of all changes at the end of the article.
%\listofchanges

\end{document}